\documentclass[aps,nofootinbib,notitlepage]{revtex4-1}

\usepackage{amsmath,amssymb}
\usepackage{bbm,graphicx}
\usepackage{hyperref}

\begin{document}

\title{Radiation-Reaction Force on a Small Charged Body to Second Order}

\author{Jordan Moxon, \'Eanna Flanagan}
\affiliation{Department of Physics, Cornell University, Ithaca, NY 14853, USA}

\begin{abstract}
In classical electrodynamics, an accelerating charged body emits radiation and
experiences a corresponding radiation-reaction force, or self force. We extend
to higher order in the total charge a previous rigorous derivation of the
electromagnetic self force in flat spacetime by Gralla, Harte, and Wald. The
method introduced by Gralla, Harte, and Wald computes the self force from the
Maxwell field equations and conservation of stress-energy in a limit where the
charge, size, and mass of the body go to zero, and does not require
regularization of a singular self field. For our higher order computation, an
adjustment of the definition of the mass of the body is necessary to avoid
including self energy from the electromagnetic field sourced by the body in the
distant past. We derive the evolution equations for the mass, spin, and
center-of-mass position of the body through second
order.  We derive, for the first time, the
second-order acceleration dependence of the evolution of the spin (self torque),
as well as a mixing between the extended body effects and the acceleration
dependent effects on the overall body motion.
\end{abstract}

\maketitle

\tableofcontents

\section{Introduction}
\subsection{Status of our understanding of self force effects}

Classical electrodynamics dictates that an accelerating charge emits
radiation. This electromagnetic radiation carries energy and momentum, so
conservation laws demand that the charge must experience a force. The force
arises from the charge interacting with its own field, and is known as the
`radiation-reaction force' or `self force'. This phenomenon was first derived by
Lorentz~\cite{Lorentz}, and later confirmed by Abraham~\cite{Abraham} followed
by Dirac~\cite{Dirac}, each expanding and generalizing the results of the prior
work.

Computing expressions for self forces is notoriously complicated, and there is
an enormous literature on this field.  The complexity arises in part because
self forces describe back-reaction: as a charge accelerates, its radiation
perturbs its motion, in turn altering the details of the radiation. Analytic
methods are tractable in the regime in which the body is small compared to the
characteristic lengthscales of the external fields. In this limit, the self
force can be expanded order by order in the charge of the body. In this paper,
we use the common nomenclature of referring to the Lorentz force as the leading
order force, the leading correction to the Lorentz force as the `first order'
self force, and so on. Our understanding of radiation reaction in flat spacetime
has been developed over most of a century
\cite{erber61,mo71,teitelboim71,spohn99}, culminating in the rigorous
treatment of Gralla, Harte, and Wald \cite{GHW}(henceforth GHW) who carefully
analyzed a limit in which the charge, size, and mass of a body go to zero. The
modern focus of the self force community is that of small masses in curved
spacetime, for which Eric Possion's review article offers a thorough
introduction~\cite{Poisson}.

The self force is of great interest to modern astrophysics. Just as a charged
particle interacts with its own field as it radiates electromagnetic waves,
gravitating systems experience self forces from the emission of gravitational
radiation. The gravitational waves produced by binary black hole inspirals and
binary neutron star inspirals have been detected by LIGO
\cite{LIGO150904,LIGO170817}, and similar binary inspirals are candidate signals
for the future space-based detector LISA.

Making full use of the data from LISA will require an improved understanding of
self force effects.  The gravitational self force to leading order in the mass
of the small body is referred to as the MiSaTaQuWa self force, and was first
derived in \cite{mino,quwa}. More recent computations have extended these
results to second order
\cite{Rosenthal,PoundConservative,PoundSummary,Detweiler,GalleyGrav1,GalleyGrav2},
and applied the self force to a gravitational inspiral, in order to compute or
numerically evaluate the worldline and the resulting gravitational
radiation. The computational strategies for evaluating worldlines and waveforms
from gravitational self force are reviewed well in
\cite{wardellStrategies,barack09}. The techniques for computing leading order,
or adiabatic, waveforms are now known. However, LISA data analysis will require
post-adiabatic waveform predictions, which in turn will also require the
subleading self force. This motivates a detailed understanding of the subleading
self force.

Previous derivations of higher-order self forces for non-gravitational fields
include those of Chad Galley \cite{Galley} and Abraham Harte
\cite{HarteExact}. Galley's derivation \cite{Galley} of the scalar self force
uses an effective field technique to derive the self force to high order for
monopolar charges. Harte has derived exact expressions for the self force of an
extended charge distribution in an external field. The relation between Harte's
results and our work is somewhat involved and is discussed in
Sec. \ref{sec:exactsolutions} below.

\subsection{The Gralla-Harte-Wald derivation method and its extension}

In this paper, we derive the subleading order electromagnetic and scalar
self forces acting on a small charged body moving in flat spacetime. The
calculation is motivated by the importance of the gravitational self force, and
is a model for the more complicated computation in the gravitational
case. Although subleading self forces have previously been computed
\cite{pound12,gralla12}, ours is the first to describe extended body effects to
subleading order. In addition to providing a model for the gravitational
self force, our calculation may have direct application to systems with
extremely strong electromagnetic fields, as discussed further below.

GHW introduce a one-parameter family of bodies with the property that as the
parameter approaches zero, the mass, charge, and spatial extent of the body
approach zero at the same rate.  By considering various moments of the
stress-energy conservation and charge conservation equations, integrated over a
small region containing the body, they derive the first-order self force, mass
evolution, and spin evolution equations.

Our calculation uses the GHW axioms with slight modifications, which are
presented in full in section \ref{sec:pointparticle}. However, we found it
necessary to modify and refine the definitions of body parameters. GHW defined
parameters such as the total mass-energy, angular momentum, and electromagnetic
multipole moments in terms of integrals over a spacelike hypersurface
perpendicular to the center of mass worldline\footnote{ As usual, there are
  ambiguities in the precise definition of center of mass worldline
  \cite{HarteExact}. These ambiguities affect the form of the equation of motion
  at subleading orders, and are associated with the choice of a spin
  supplementary condition. See Section \ref{sec:DHparams} below.}. At second
order, these definitions are problematic, and we replace them with body
parameter definitions in terms of integrals over the future null cones of points
on the center of mass worldline. With these definitions, the body parameters at
a given time depend only on the body's stress-energy and charge distribution at
times within a light crossing time, not on the stress-energy or charge
distribution in the distant past. This is because, in flat spacetime, the field
at every point depends only on sources on that point's past lightcone.

\subsection{Discussion of results - applications in physical systems} \label{sec:applications}

Our results for the second order evolution of the body's worldline, mass, and
spin are given in Eqs. (\ref{eq:secondorderstart}) -
(\ref{eq:diquadMdotSecond}). They contain three types of terms: coupling of
electromagnetic moments to the external field, self force terms that do not
depend on the higher electromagnetic moments, and terms which describe a mixing
between self-field and extended body effects. Our spin evolution equation
contains a self-torque, which was not seen previously at lower orders. Our
results also satisfy a consistency check obtained by comparing with some
non-perturbative results of Harte \cite{HarteExact}.

As an illustrative special case, consider a body with vanishing spin, electromagnetic dipole, and quadrupole, moving in an external electromagnetic field $F^{(\text{ext}) \mu \nu}$. The acceleration of the body can be written as [c.f. Eq. (\ref{eq:ppselfForce}) below], in units with $c=1$,
\begin{align}
  a^\mu =& \kappa F^{(\text{ext}) \mu \lambda} u_\lambda + q\left\{ \frac{2}{3} \kappa^2 D_\tau F^{(\text{ext}) \mu \lambda} u_\lambda + \frac{2}{3} \kappa^3 \mathcal{P}^\mu{}_\nu F^{(\text{ext}) \nu \lambda} F^{(\text{ext})}{}_{\lambda \sigma} u^\sigma\right\} \notag\\
  & + q^2 \bigg\{ \frac{4}{9} \kappa^3 D_\tau{}^2 F^{(\text{ext}) \mu \lambda} u_\lambda + \frac{8}{9} \kappa^4 \mathcal{P}^\mu{}_\kappa F^{(\text{ext}) \kappa \lambda} F^{(\text{ext})}{}_{\lambda \sigma} u^\sigma\notag \\
  &\hspace{1cm} +\frac{4}{9} \kappa^4 \mathcal{P}^\mu{}_\kappa F^{(\text{ext})\kappa \lambda} D_\tau F^{(\text{ext})}_{\lambda \sigma} u^\sigma + \frac{4}{9}\kappa^5 \mathcal{P}^\mu{}_\kappa F^{(\text{ext}) \kappa \rho} \mathcal{P}_{\rho \lambda} F^{(\text{ext}) \lambda \sigma} F^{(\text{ext})}{}_{\sigma \omega} u^\omega \bigg\} + \mathcal{O}(q^3).
\end{align}
Here $u^\mu$ is the 4-velocity of the body, $a^\mu$ the 4-acceleration, $D_\tau
\equiv u^\mu \nabla_\mu$, and $\mathcal{P}^\mu{}_\nu = \delta^\mu{}_\nu + u^\mu
u_\nu$ is the projection tensor. Also, $q$ is the charge and $\kappa = q/m$ is
the charge to mass ratio. The right hand side consists of an expansion in $q$ at
fixed $\kappa$. The first term is the Lorentz force law, the second term is the
reduced-order (see Sec. \ref{sec:preamble} below) form of the
Abraham-Lorentz-Dirac equation, and the third term is our new result.

\begin{figure*}
  \includegraphics[width=.85\textwidth]{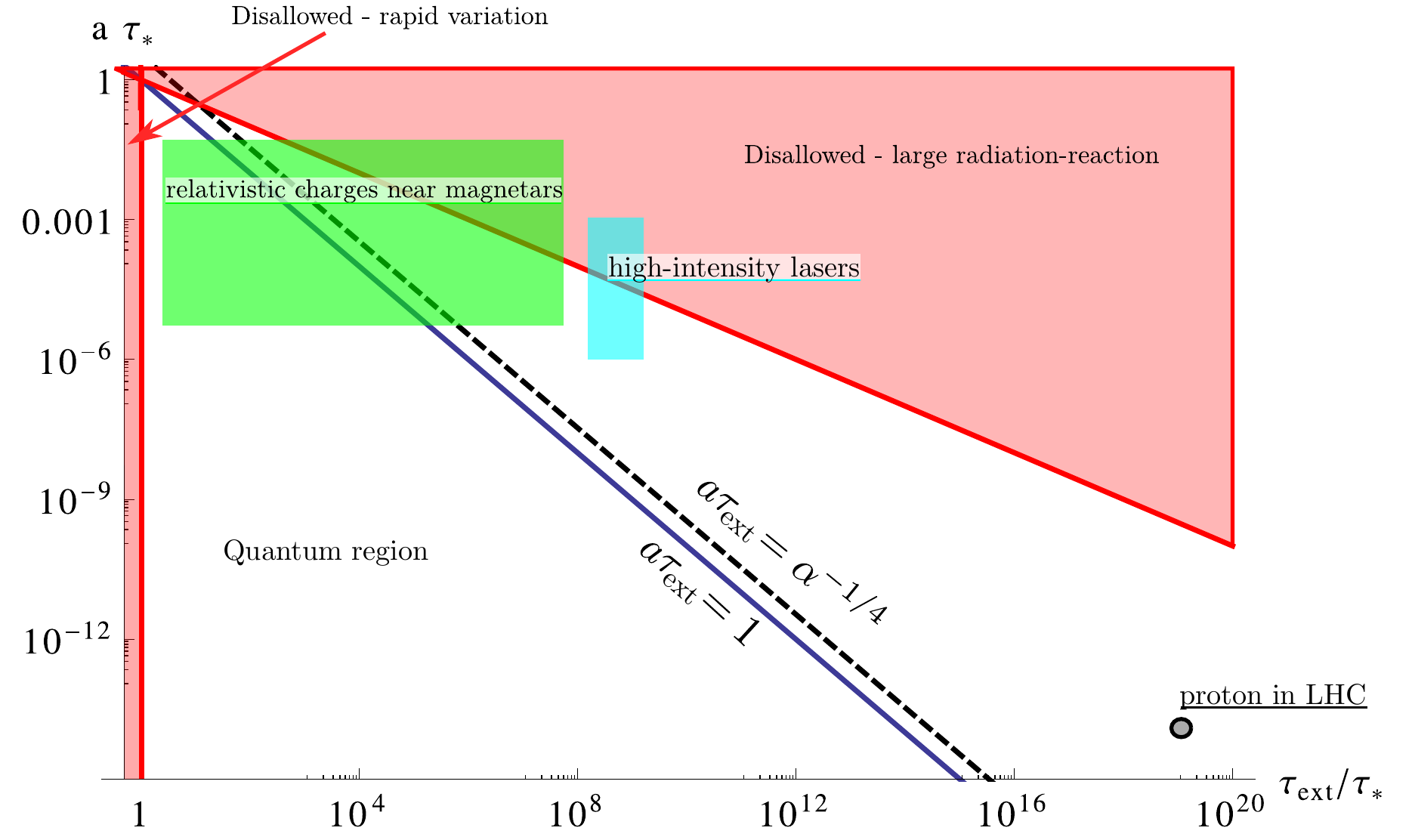}
  \caption{\label{fig:scalesPlot} An illustration of the parameter space for radiation reaction for relativistic particles. The horizontal axis is the ratio $\tau_{\text{ext}}/\tau_*$, where $\tau_{\text{ext}}$ is the timescale over which the external field is varying, as measured in the instantaneous rest frame of the particle, and $\tau_* = q^2/m$, where $q$ is the charge and $m$ the mass of the particle. The vertical axis is $a \tau_*$, where $a$ is the acceleration due to the external field. The motion is relativistic in the region $a\tau_{\text{ext}}\gg 1$, in the upper right hand of the figure. Radiation reaction effects are large in the shaded regions where $\tau_*/\tau_{\text{ext}}\gtrsim 1$ or $a^2 \tau_{\text{ext}} \tau_* \gtrsim 1$. These regions lie outside of the domain of validity of our analysis, and the second order self force is negligible except near the boundaries of these regions. In the region below and to the left of the dashed line, the radiation from the particle is not in a classical regime and our analysis does not apply. We show the location of protons  in the Large Hadron Collider, protons in very high intensity lasers, and electrons in the high magnetic fields of magnetars. }
\end{figure*}

We now turn to a discussion of the domain of validity of our results. Consider a
charged body of mass $m$, and charge $q$, moving in an external field that
imparts a characteristic acceleration $a$, as measured in the body's
instantaneous rest-frame. Suppose also that the field varies on some timescale
or lengthscale $\tau_{\text{ext}}$, again as measured in the body's
instantaneous rest-frame. Then there are a number of conditions that must be
satisfied for our analysis to be valid:
\begin{itemize}
\item \emph{Small multipole couplings}: If the condition
  \begin{equation}
    \mathcal{R} \ll \tau_{\text{ext}}, \label{eq:smallcond}
  \end{equation}
  is satisfied then the leading order couplings (dipole, quadrupole, and so on) will dominate.
\item \emph{Weak radiation reaction}: The energy radiated in a dynamical time must be small compared to the change in the body's energy due to conservative effects. If this is violated then our derivation is no longer valid. In the non-relativistic region $a\tau_{\text{ext}}\ll 1$ this requires
  \begin{equation}
    \frac{\tau_*}{\tau_{\text{ext}}} \ll 1 \label{eq:rrcond1}
  \end{equation}
  where $\tau_* = q^2/m$. In the relativistic regime $a \tau_{\text{ext}}\gg 1$, the condition is instead
  \begin{equation}
    a^2 \tau_{\text{ext}} \tau_* \ll 1. \label{eq:rrcond2}
  \end{equation}
\item \emph{Classical radiation regime}: The energy radiated in a dynamical time must be large compared to the energy radiated per quantum, so that many quanta are emitted in a dynamical time. In the non-relativistic regime $a\tau_{\text{ext}} \ll 1$ the corresponding requirement is
  \begin{equation}
    a \tau_{\text{ext}} \gg \alpha^{-1/2},
  \end{equation}
  where $\alpha=q^2/\hslash$, and the relativistic regime $a\tau_{\text{ext}}\gg1$ it is
  \begin{equation}
    a \tau_{\text{ext}} \gg \alpha^{-1/4}. \label{eq:quantumrr}
  \end{equation}
\end{itemize}
For elementary particles typically $\alpha \ll 1$ while for macroscopic charged bodies $\alpha \gg 1$.

Our derivation method employs a certain limiting procedure which automatically
enforces the conditions (\ref{eq:smallcond}),(\ref{eq:rrcond1}), and
(\ref{eq:rrcond2}). The two dimensional parameter space of acceleration $a$ and
external timescale $\tau_{\text{ext}}$ is illustrated in Fig
\ref{fig:scalesPlot}. The solid line $a\tau_{\text{ext}} =1$ is the boundary between
non-relativistic and relativistic motion; the lower left region is
non-relativistic while the upper right is relativistic. The shaded regions on
the left and at the top correspond to strong radiation reaction and lie outside
our domain of validity, by (\ref{eq:rrcond1}) and (\ref{eq:rrcond2}). Our second
order self force will be significant only near these boundaries. The region to
the left of the dashed line is disallowed since the radiation is not classical,
by (\ref{eq:quantumrr}) (assuming an elementary particle so that $\alpha \ll
1$). Also shown on the plot are some illustrative examples:

\begin{itemize}
\item A proton at the Large Hadron Collider, for which $a\sim 3\cdot 10^{12}\,\text{s}^{-1}$, $\tau_{\text{ext}}\sim 1.4\cdot 10^{-8}\,\text{s}$, $\tau_* \sim 6\cdot 10^{-27}\,\text{s}$. In this case we have $a^2 \tau_{\text{ext}} \tau_* \sim 10^{-9}$, so higher order radiation reaction effects are negligible. Lead ions in the LHC experience a similar acceleration, and have a $\tau_*$ almost two orders of magnitude larger, $\tau_* \sim 2 \cdot 10^{-25}\,\text{s}$, so the scale of effect is $a^2 \tau_{\text{ext}} \tau_* \sim 10^{-8}$. 
\item For high-intensity laser systems with intensities in the range $10^{19}\,\text{W}/\text{cm}^2 - 10^{22}\,\text{W}/\text{cm}^2$ \cite{KumarRRplasma,BerezhianiRRplasma,ChenRRplasma}, the acceleration scale for a proton is then in the range $a\sim 10^{17}\,\text{s}^{-1} - 10^{21}\,\text{s}^{-1}$, and using $\tau_{\text{ext}}\sim 10^{-16}\,\text{s}$ and $\tau_* \sim 6 \cdot 10^{-23}\,\text{s}$ gives $a^2 \tau_{\text{ext}} \tau_*$ in the range $10^{-8}$-$10^{0}$. At the upper end of this range, second order radiation reaction effects could become significant. \cite{krueger}
  \item Turning to astrophysics, the magnetic fields near certain neutron stars, referred to as ``magnetars'', can be extremely large, $B\sim 10^{8}-10^{11} \text{T}$. At the high end of this range, higher order self force effects could easily become large even for slowly moving particles.
\end{itemize}

\section{Motion of a finite body coupled to an external field} \label{sec:motion}

In this section, we consider a finite extended body moving in an external field
in flat spacetime. We will review the governing equation, the non-perturbative
definition of the body parameters.  In the following sections we will review the
non-perturbative equations of motion for the body moments, and specialize to the
limit of a small body to obtain explicit results.

\subsection{Governing equations} \label{sec:GovEq}

The system we are considering is a finite, extended, charged body coupled to an
external field in flat spacetime.  The extended body is described by a matter
stress-energy tensor $T^{\mu \nu}_M$, which we assume is smooth and which
vanishes outside a world tube of compact spatial support. We will consider both
electromagnetic and scalar self forces.

The coupling to either type of field is governed by the body's charge, which is
described by a charge current density $j^\mu$ such that $\nabla_\mu j^\mu = 0$
(electromagnetic case), or a scalar charge density $\rho$ (scalar case). We
assume that the charge current or density functions are also smooth and of
compact spatial support. These fields obey the standard inhomogeneous wave
equations for the respective type of field:
\begin{subequations} \label{eq:maxwell}
\begin{align} 
\qquad \qquad \nabla_{[\mu} F_{\lambda \sigma]}& = 0, \\ \qquad \qquad
\nabla^\mu F_{\mu \nu}& = 4 \pi j_\nu \qquad (\text{E\&M case}),
\end{align}
\end{subequations}
and 
\begin{equation} \label{eq:scalarfield}
\qquad \qquad \nabla_\mu \nabla^\mu \Phi = -4 \pi \rho \qquad (\text{scalar
  case}).
\end{equation}

The total stress-energy tensor $T_{\mu \nu}$ is given by the sum of the matter
contribution $T_{M \, \mu \nu}$ and the field contribution $T_{F\, \mu
  \nu}$. This stress energy contribution for the electromagnetic field is
\begin{equation} \label{eq:fieldSE}
\qquad \qquad 4\pi T_{F \, \mu \nu} = F_{\mu \lambda} F^{\lambda}{}_{\nu}
- \frac{1}{4} g_{\mu \nu} F_{\sigma \lambda} F^{\sigma \lambda},
\end{equation}
or, for the scalar field, is 
\begin{equation} \label{eq:scalarfieldSE}
\qquad \qquad 4\pi T_{F\,\mu \nu} = \nabla_\mu \Phi \nabla_\nu \Phi -
\frac{1}{2} g_{\mu \nu} \nabla_\lambda \Phi \nabla^\lambda \Phi.
\end{equation}
We assume that this total stress-energy is conserved:
\begin{equation} \label{eq:secons}
  \nabla_\mu \left(T_{\text{M}}^{\mu \nu} + T_{\text{F}}^{\mu \nu}\right) = 0.
\end{equation}

We choose to divide the field into an external field $F^{(\text{ext}) \mu \nu}$
(Scalar: $\Phi^{(\text{ext})}$), and a self field $F^{(\text{self}) \mu
  \nu}$ (Scalar: $\Phi^{(\text{self})}$) which is the retarded solution to
the field equations (\ref{eq:maxwell}) or (\ref{eq:scalarfield}) with the given
source. The external field may be expressed as, for the electromagnetic case,
\begin{equation} \label{eq:fieldSplit}
\qquad\qquad F^{(\text{ext})}{}_{\mu \nu} = F_{\mu \nu} -
F^{(\text{self})}{}_{\mu \nu},
\end{equation}
or, for the scalar case,
\begin{equation} \label{eq:scalarfieldSplit}
\quad \qquad \qquad \Phi^{(\text{ext})}{}_{} = \Phi_{} -
\Phi^{(\text{self})}{}_{} .
\end{equation}
Inserting the decompositions (\ref{eq:fieldSplit}),(\ref{eq:scalarfieldSplit}) into the quadratic expressions (\ref{eq:fieldSE}),(\ref{eq:scalarfieldSE}) for the field stress energy tensor, we find following GHW that the field stress energy can be expressed as the sum of three terms:
\begin{equation} \label{eq:SEselfcrossext}
  T_{\text{F}}^{\mu \nu} = T^{\mu \nu}_{(\text{self})} + T^{\mu
    \nu}_{(\text{cross})} + T^{\mu \nu}_{(\text{ext})} .
\end{equation}
Here $T^{\mu \nu}_{(\text{self})}$ is quadratic in the self field, $T^{\mu \nu}_{(\text{ext})}$ is quadratic in the external field, and $T^{\mu \nu}_{(\text{cross})}$ is a cross term which depends on both the self field and the external field.

In the following subsection we will discuss the definition of body parameters such as mass, momentum, and spin. For those definitions, we will use the sum of the matter and self stress energy tensors,
\begin{equation} \label{eq:bodySE}
  T^{\mu \nu} = T_{\text{M}}^{\mu \nu} + T_{(\text{self})}^{\mu \nu},
\end{equation}
excluding the cross and external contribution, following GHW.  The conservation
of stress-energy (\ref{eq:secons}) can be rewritten in terms of this quantity as:
\begin{subequations}\label{eq:seEqs}
\begin{align} \label{eq:seEqsMaxwell}
\nabla_\mu T^{\mu \nu} = F^{(\text{ext})\nu \mu} j_\mu \qquad \text{(E\&M case)},
\\ \nabla_\mu T^{\mu \nu} = \Phi^{(\text{ext});\nu} \rho \qquad \text{(scalar case)}.
\end{align}
\end{subequations}
The motivation for choosing the definition (\ref{eq:bodySE}) for the body
parameter definitions is that in the limit when the body becomes small, the fields $T^{\mu
  \nu}$, $j^{\mu}$, and $\rho$ vary over the small body lengthscale, while the
external fields $F^{(\text{ext}) \mu \nu}$ and $\Phi^{(\text{ext});\mu}$ vary
only on a longer lengthscale set by the external field.

The only equations that are needed for our derivation of the self force are the field equations (\ref{eq:maxwell}) and (\ref{eq:scalarfield}), the stress energy conservation equation in the form (\ref{eq:seEqs}), and the definition of the self-field as the retarded field.

\subsection{Non-perturbative definition of body parameters: the Dixon-Harte formalism} \label{sec:DHparams}

We now turn to a discussion of the definition of body parameters for a finite
body, including the body's mass, momentum, spin, and choice of representative
worldline.

For a conserved stress energy tensor $T^{\mu \nu}$ in flat spacetime of compact spatial
support, there is a natural choice of momentum and spin, namely
\begin{subequations}\label{eq:Minspinmom}
\begin{align}
  P_{(\text{Isolated})}^\mu =& \int_\Sigma T^{\mu \nu} d \Sigma_\nu,\\
  S_{(\text{Isolated})}^{\mu \nu}(z^\mu) =& 2\int_{\Sigma} (x - z)^{[\mu} T^{\nu] \lambda} d \Sigma_\lambda,
\end{align}
\end{subequations}
where $\Sigma$ is any spacelike hypersurface. The center of mass worldline is
then the set of points $z^\mu$ which satisfy
\begin{equation} \label{eq:Minspinsup}
  S_{(\text{Isolated})}^{\mu \nu}(z^\mu) P_{(\text{Isolated}) \nu} =0.
\end{equation}
Equation (\ref{eq:Minspinsup}) is known as a spin supplementary condition, and
generalizations of this condition will be discussed below.

However, this treatment is not applicable to our present context for two
reasons:
\begin{itemize}
  \item First, the stress-energy tensor (\ref{eq:bodySE}) that we wish to use in the
definitions is not conserved, instead there is a forcing term from the external
field on the right hand side of Eqs. (\ref{eq:seEqs}). Hence, the expressions
(\ref{eq:Minspinmom}) will no longer be independent of the choice of
hypersurface $\Sigma$, and a specific choice of hypersurface $\Sigma$ will be
required. This will be discussed further below.
\item Second, the stress energy term (\ref{eq:bodySE}) that we will use does not have
compact spatial support, due to the self field contribution. Hence, there is no
guarantee that the expressions (\ref{eq:Minspinmom}) are convergent and well
defined. The convergence of these integrals is discussed further below.
\end{itemize}

There exists a  general, fully non-perturbative set of definitions
of worldlines, electromagnetic moments, and stress-energy moments of an extended
body. These definitions were introduced by Dixon \cite{dixon1,dixon2} in the
context of curved spacetime, and extended by Harte \cite{HarteExact}. We follow
the Dixon-Harte framework and definitions, with some modifications that we
discuss below. The remainder of this section reviews those aspects of the
Dixon-Harte framework that are most important for our derivation.

\begin{figure*}[t]
  \includegraphics[width=.8\textwidth]{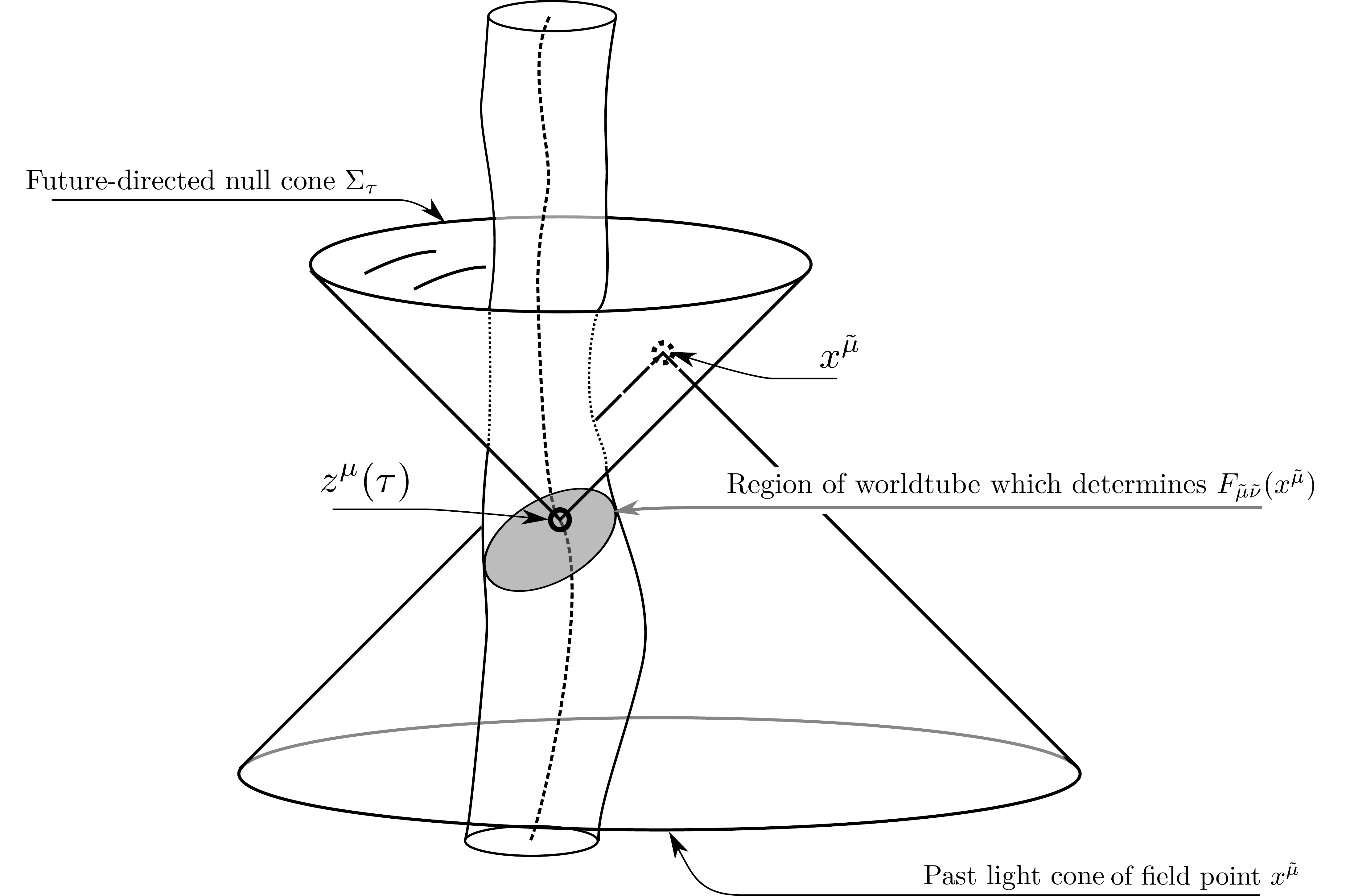}
  \caption{An illustration of our definitions of total momentum and spin of an extended body. The body is confined to the world tube shown, but is coupled to a long range field  (scalar or electromagnetic) that extends beyond the worldtube. Given a representative worldline $z^\mu(\tau)$, shown as a dashed line, we define momentum and spin by integrating over future null cones $\Sigma_\tau$ of points on the worldline. The field stress energy tensor at a point $x^{\tilde{\mu}}$ on such a null cone will depend on the sources in the intersection of its past lightcone with the worldtube, shaded in gray. This region is confined to within the region of the worldtube consisting of times $\tau^\prime$ with $|\tau - \tau^\prime|$ smaller than a light-crossing time. }\label{fig:cartoon}
\end{figure*}

Before discussing the definitions of body parameters, we review the covariant
bitensor formalism \cite{Poisson}. We work in flat spacetime, but we will be
using non-Lorentzian coordinates. We will denote by $x^{\tilde{\mu}}$ a field
point off the worldline, and we use tilded indices for tensors at such
points. We will denote by $z^\mu(\tau)$ a point on the worldline (figure
\ref{fig:cartoon}), and use normal (untilded) indices for the tensors at such
points. General bitensors are functions of both $z^\mu$ and $x^{\tilde{\mu}}$,
and can have one or more indices of either type.

An important set of bitensors are Synge's worldfunction $\sigma(x,z)$ and its
derivatives. Synge's worldfunction is defined only for pairs of points that are
sufficiently close that there exists a unique geodesic that joins them. For this
unique geodesic, $\sigma(x,z)$ measures the half geodesic distance squared
between the two points. It is negative for timelike separated points, positive
for spacelike separated points, and zero for null-related points. The first
covariant derivative of Synge's worldfunction can be used to define a covariant
version of a position vector $\sigma_{\mu}(x,z) \equiv \nabla_\mu \sigma(x,z)$,
where the derivative is with respect to $z$. We will also find useful the second
derivatives, $\sigma^\mu{}_\lambda(x,z) \equiv \nabla_\lambda \nabla^\mu \sigma$
and $\sigma^{\tilde{\mu}}{}_\lambda \equiv \nabla^{\tilde{\mu}} \nabla_\lambda
\sigma$.

In the Dixon-Harte framework, one chooses a worldline $z^\alpha(\tau)$ for the
body, where $\tau$ is a parameter that need not be proper time, and a choice of
a unit vector $n^\alpha(\tau)$ along the worldline with $n_\alpha (d/d\tau)^\alpha
= -1$. The formalism supplies conditions that eventually determine the worldline
and parameterization. Given these choices, one defines a foliation of spacetime
by hypersurfaces $\Sigma_\tau$ as follows.  Each hypersurface is labeled by the
parameter $\tau$ at which it intersects the worldline, so $z^\alpha(\tau) \in
\Sigma_\tau$, and is generated by geodesics starting on the worldline that are
orthogonal to $n^\alpha$.

The Dixon-Harte definitions of the momentum and spin of an extended body are
\begin{subequations}
  \begin{align}
    P_{\mu}^{\text{D}}(\tau) =& \int_{\Sigma_\tau} d \Sigma_{\tilde{\mu}}(x)
    T_M^{\tilde{\mu} \tilde{\nu}}(x) K_{\tilde{\nu} \mu}(x,z_\tau), \\
    S_{\mu \nu}^{\text{D}}(\tau) =& 2 \int_{\Sigma_\tau} d\Sigma_{\tilde{\mu}}(x)
    T_M^{\tilde{\mu} \tilde{\nu}}(x) H_{\tilde{\nu} [\mu}(x,z_\tau)
      \sigma_{\nu]}(x,z_\tau),
  \end{align}
\end{subequations}~where~\begin{subequations}
  \begin{align}
    H^{\tilde{\mu}}{}_\nu =& -\left(\sigma^\nu{}_{\tilde{\mu}}\right)^{-1}, \\
    K^{\tilde{\mu}}{}_\nu =& H^{\tilde{\mu}}{}_\lambda \sigma^\lambda{}_\nu.
  \end{align}
\end{subequations}

In flat spacetime, these definitions reduce to:

\begin{subequations} \label{eq:DHmomSpinFlatDef}
  \begin{align} 
    P_D^{\mu}(\tau) =& \int_{\Sigma_\tau} d \Sigma_{\tilde{\mu}}(x^{\tilde{\lambda}})
    T_M^{\tilde{\mu} \tilde{\nu}}(x^{\tilde{\lambda}}) g_{\tilde{\nu}}{}^{\mu}(x^{\tilde{\lambda}},z^\lambda(\tau)),  \\
    S_D^{\mu \nu}(\tau) =& 2 \int_{\Sigma_\tau} d\Sigma_{\tilde{\mu}}(x^{\tilde{\lambda}})
    T_M^{\tilde{\mu} \tilde{\nu}}(x^{\tilde{\lambda}}) g^{}_{\tilde{\nu}}{}^{[\mu}(x^{\tilde{\lambda}},z^\lambda)
      \sigma^{\nu]}(x^{\tilde{\lambda}},z^\lambda), 
  \end{align}
\end{subequations}
where $g_{\mu \tilde{\nu}} \equiv - \sigma_{\mu \tilde{\nu}}$ is the parallel
propagator bitensor in flat spacetime. 

We modify the Dixon-Harte framework in the following ways.
\begin{itemize}
\item We specialize the parameter $\tau$ to be the proper time.
\item We dispense with the unit vector $n^\alpha(\tau)$.
\item We use the stress energy tensor $T^{\mu \nu}$ of Eq. (\ref{eq:bodySE})
  instead of the matter stress energy tensor $T_M^{\mu \nu}$.
\item We use null hypersurfaces $\Sigma_\tau$ that are generated by the set of
future null geodesics starting at worldline point $z^\alpha(\tau)$.
This family of null hypersurfaces foliates the convex normal neighborhood of the
worldline, which covers the entire manifold for the flat spacetime case we
consider in this paper.
\end{itemize}
Our definitions are then
\begin{subequations} \label{eq:momSpinFlatDef}
  \begin{align} 
    P_B^{\mu}(\tau) =& \int_{\Sigma_\tau} d \Sigma_{\tilde{\mu}}(x^{\tilde{\lambda}})
    T^{\tilde{\mu} \tilde{\nu}}(x^{\tilde{\lambda}}) g_{\tilde{\nu}}{}^{\mu}(x^{\tilde{\lambda}},z^\lambda(\tau)), \label{eq:momFlatDef} \\
    S_B^{\mu \nu}(\tau) =& 2 \int_{\Sigma_\tau} d\Sigma_{\tilde{\mu}}(x^{\tilde{\lambda}})
    T^{\tilde{\mu} \tilde{\nu}}(x^{\tilde{\lambda}}) g^{}_{\tilde{\nu}}{}^{[\mu}(x^{\tilde{\lambda}},z^\lambda)
      \sigma^{\nu]}(x^{\tilde{\lambda}},z^\lambda), \label{eq:spinFlatDef}
  \end{align}
\end{subequations}
 Here the subscript $B$ denotes ``bare''; these definitions will be replaced by
 renormalized momentum and spin in Sec. \ref{sec:restricted} below.

The motivations for our choice of foliation of future null cones are as follows.
The integrals (\ref{eq:Minspinmom}) contain a contribution from the stress
energy tensor of the self field from Eq. (\ref{eq:bodySE}). That self field,
evaluated at a point $x$ on the hypersurface $\Sigma_\tau$ over which one
integrates, in turn depends on the body's charge distribution on the past light
cone of $x$. When one uses a spacelike hypersurface $\Sigma_\tau$, the
dependence on the body's charge distribution extends into the distant past, as
one takes $x$ further and further out on the spacelike hypersurface. By
contrast, for a future null cone, $\Sigma_\tau$, the dependence on the body's
charge distribution is limited to times within a light-crossing time of $\tau$,
as illustrated in figure (\ref{fig:cartoon}). In addition, we show in Appendix
\ref{app:sescaling} that the integrals (\ref{eq:momSpinFlatDef}) are well
defined and finite when the hypersurfaces $\Sigma_\tau$ are chosen to be future
null cones.

There are three choices we have alluded to in the above definition of momentum
and spin: the worldline $z(\tau)$ (which is fixed by the spin supplementary
condition), the choice (\ref{eq:bodySE}) of body stress-energy tensor, and the
choice of the hypersurface of integration. As we have argued, not all choices
give rise to physically acceptable definitions. Within those that do there is
considerable freedom. This freedom corresponds to different ways of describing a
given dynamical system.  Different choices will give rise to different forms of
the laws of motion, but will not change any physical predictions.

We also define the bare rest mass $m_B$ by
\begin{equation} \label{eq:restmass}
  m_B^2 \equiv - P_B^\mu P_{B\,\mu}.
\end{equation}
We define the 4-velocity in the usual way as $u^\mu(\tau) = dz^\mu/d\tau$, with
$u^\mu u_\mu = -1$,  and note that
\begin{equation}
 P_B^\mu \ne m_B u^\mu,
\end{equation}
beyond leading order.

The definitions (\ref{eq:momSpinFlatDef}) are valid for any choice of worldline
$z_\tau$. To pick out a unique worldline one must specify a spin supplementary
condition \cite{dixon1,dixon2}, which takes the generic form
\begin{equation} \label{eq:spinsupgen}
 S_B^{\mu \nu}(\tau) \omega_\nu = 0,
\end{equation}
where $\omega_\nu$ is some vector field defined on the worldline. Such a spin
supplementary condition defines a center of mass worldline \cite{kyrian2007}
\cite{costa2012}. Our the spin supplementary condition is defined in terms of a
renormalized spin $S^{\mu \nu}$, which we define in
Eq. (\ref{eq:restrictedSpin}) below. Our spin supplementary condition is
\begin{equation} \label{eq:spinsup}
   S^{\mu \nu} u_\nu = 0,
\end{equation}
which reduces at leading order in the size and mass of the body to the condition
(\ref{eq:spinsupgen}) with $\omega_\nu = u_\nu$.

\subsection{Electromagnetic multipole moments} \label{sec:moments}

We now turn to a discussion of electromagnetic multipole moments. We define the
total (conserved) bare charge $q_B$, charge moment $\mathcal{J}^{\mu}_B$, dipole
$Q_B^{\mu \nu}$, and quadrupole $Q_B^{\mu \nu \rho}$ of the body to be
\begin{subequations} \label {eq:baremult}
\begin{align}
  q_B(\tau) = q_B =& \int_{\Sigma_\tau} d \Sigma_{\tilde{\nu}} j^{\tilde{\nu}}, \label{eq:bareCharge}
  \\
  \mathcal{J}_B^{\mu}(\tau) =& \int_{\Sigma_\tau} d\Sigma_{\tilde{\nu}}
  g^{\tilde{\nu}}{}_{\lambda} u^{\lambda} j^{\tilde{\mu}}
  g_{\tilde{\mu}}{}^{\mu} ,\label{eq:bareChargeMoment} \\
  Q_B^{\mu \nu}(\tau) =& -\int_{\Sigma_\tau} d\Sigma_{\tilde{\nu}}
  g^{\tilde{\nu}}{}_{\lambda} u^{\lambda} j^{\tilde{\mu}}
  g_{\tilde{\mu}}{}^{\mu}\sigma^{\nu}, \label{eq:bareDipole}\\
  Q_B^{\mu \nu \rho}(\tau) =& \int_{\Sigma_\tau} d\Sigma_{\tilde{\nu}}
  g^{\tilde{\nu}}{}_{\lambda} u^{\lambda} j^{\tilde{\mu}}
  g_{\tilde{\mu}}{}^{\mu} \sigma^{\nu} \sigma^{\rho}. \label{eq:bareQuadrupole}
\end{align}
\end{subequations}
 In these expressions, the arguments of all the bitensors
 $g^{\tilde{\nu}}{}_\lambda$, $\sigma^\nu$, etc. are $(x,z(\tau))$, while the
 argument of $j^{\tilde{\mu}}$ is $(x)$. The definition (\ref{eq:bareDipole})
 has a minus sign due to the properties of Synge's worldfunction
 ($g_\mu{}^{\tilde{\nu}} \sigma^\mu = -\sigma^{\tilde{\nu}}$).

For the Dixon moments \cite{dixon2} defined in terms of a spacelike hypersurface
generated by geodesics orthogonal to $n^\mu(\tau)$, the bitensor
$\sigma^\mu(x,z(\tau))$ is orthogonal to $n^\mu(\tau)$ for all $x$ in
$\Sigma_\tau$, and hence all of the charge moments are orthogonal to $n^\mu$ in
all indices following the first index:
\begin{equation}
  Q_D^{\mu \nu} n_\mu = Q_D^{\mu \nu \rho} n_\mu = Q_D^{\mu \nu \rho} n_\rho = 0.
\end{equation}
Since we integrate over future-directed null cones, there is no such
orthogonality condition for our moments (\ref{eq:baremult}). In addition, our
dipole (\ref{eq:bareDipole}) contains both a symmetric and an antisymmetric
part, unlike the case for the standard definition which includes an explicit
antisymmetrization.

The number of independent components of the electromagnetic dipole
(\ref{eq:bareDipole}) and quadrupole (\ref{eq:bareQuadrupole}) are nominally 16
and 40, respectively. When charge conservation is imposed in
Sec. \ref{sec:computationSteps}, we shall see that these reduce to 10 and
22. However, these are still larger than the number of degrees of freedom for
the standard definitions of the electromagnetic dipole and quadrupole, which are
6 and 14. Our bare electromagnetic moments (\ref{eq:baremult}) are convenient
for our derivation in Sec. \ref{sec:Computation}. However, we shall express our
final results for the equations of motion in terms of a set of renormalized,
projected moments, defined in Sec. \ref{sec:restricted}, which have the standard
number of degrees of freedom.

\subsection{Scalar multipole moments}

For the scalar case, we define an analogous set of bare moments, based on
integrals over the scalar source $\rho$,
\begin{subequations} \label {eq:baremultScalar}
\begin{align}
  q_{S B}(\tau) =& \int_{\Sigma_\tau} d \Sigma_{\tilde{\nu}} u^{\tilde{\nu}}
  \rho, \label{eq:bareChargeScalar}\\ Q_{S B}^{\mu}(\tau) =& -\int_{\Sigma_\tau}
  d\Sigma_{\tilde{\nu}} g^{\tilde{\nu}}{}_{\lambda} u^{\lambda} \rho
  \sigma^{\mu}, \label{eq:bareDipoleScalar}\\ Q^{\mu \nu}_{S B}(\tau) =&
  \int_{\Sigma_\tau} d\Sigma_{\tilde{\nu}} g^{\tilde{\nu}}{}_{\lambda}
  u^{\lambda} \rho \sigma^{\mu} \sigma^{\nu} \label{eq:bareQuadrupoleScalar}.
\end{align}
\end{subequations}
All other details regarding the absence of an orthogonality condition, and the
comparison to standard multipoles are similar to those for the electromagnetic
multipoles. Here the subscript $S$ denotes ``scalar'' and $B$ denotes ``bare''.

The multipole moments (\ref{eq:baremult}) and (\ref{eq:baremultScalar}) that we
are defining are non-standard. However, they contain the same information as
standard multipole moments which are defined in terms of integrals over
spacelike hypersurfaces. Some insight into the relation between the two sets of
moments can be obtained by considering the leading order expansion for $\Phi$ in
terms of its source $\rho$ in a Lorentz frame $(t,x^i)$:
\begin{equation}
  \Phi(t,r,n^i) = \frac{1}{r} \int d^3 y \rho(t-r + {\bf n}\cdot {\bf y}, {\bf y}) + \mathcal{O}\left(\frac{1}{r^2}\right),
\end{equation}
where
\begin{equation}
  r= |{\bf x}| \hspace{1cm} \text{and} \hspace{1cm} n^i = \frac{x^i}{r}.
\end{equation}
Taylor expanding the density about the retarded time $t-r$ gives the usual multipole expression
\begin{align} \label{eq:standardexpansionscalar}
  \Phi(t,r,n^i) = \frac{1}{r} \sum_{k=0}^\infty \bigg[\frac{1}{k!} n^{i_1} \dots n^{i_k} \int d^3 y y^{i_1} \dots y^{i_k} \rho^{(k)} (t-r,{\bf y})\bigg] + \mathcal{O}\left(\frac{1}{r^2}\right),
\end{align}
where $\rho^{(k)}$ denotes the $k^{\text{th}}$ time derivative. Taylor expanding instead about $r-t+y$ yields
\begin{align} \label{eq:scalarfieldexpansion}
  \Phi(t,r,n^i) = \frac{1}{r} \sum_{k=0}^{\infty}&\left[ \frac{1}{k!} \int d^3 y(n^i y^i - y)^k \rho^{(k)}(t-r + y,{\bf y})\right]  + \mathcal{O}\left(\frac{1}{r^2}\right),
\end{align}
which now involves integral over the future null cones. The integrals that appear in (\ref{eq:scalarfieldexpansion}) are precisely time derivatives of our nonstandard multipoles (\ref{eq:baremultScalar}) 

\section{Non-perturbative equations of motion} \label{sec:exactsolutions} \label{sec:nonperturbative}

This paper focuses primarily on a perturbative expansion of the self force. It
is informative, though, to consider the extent to which exact computations can
be used to determine radiation-reaction effects. In this section, we derive an
exact law of motion for extended bodies, which is used indirectly in our
derivation in the remainder of the paper. Our exact law is a modification of an
exact law of motion due to Harte \cite{dixon1,HarteExact}, which we review. We
use Harte's result to perform a consistency check of our results in
Sec.\ref{sec:Computation} below.

\subsection{Equation of motion for bare momentum}

First, we define a generalized momentum $P_\tau(\vec{\xi})$ as a linear map on vector fields
$\xi^{\tilde{\mu}}$ via
\begin{equation} \label{eq:harteMomentum}
  P_\tau(\vec{\xi}) = \int_{\Sigma_\tau} T^{\tilde{\mu}\tilde{ \nu}} \xi_{\tilde{\mu}}
  d\Sigma_{\tilde{\nu}}.
\end{equation}
Here, as before, we choose the surface $\Sigma_\tau$ of integration to be
future-directed null cones. When we specialize $\vec{\xi}$ to be a Killing
vector field $\xi_{\tilde{\mu}} = g_{\tilde{\mu}}{}^\mu$ or $\xi_{\tilde{\mu}} =
2 g_{\tilde{\mu}}{}^{[\sigma} \sigma^{\nu]}$, the resulting quantities
(\ref{eq:harteMomentum}) yields the definitions (\ref{eq:momSpinFlatDef}) of
linear momentum and spin \cite{HarteExact}.

To compute the time derivative of this generalized momentum, we use the general
identity \cite{HarteExact}
\begin{equation} \label{eq:surfaceTrick}
  \frac{d}{d\tau} \int_{\Sigma_\tau} v^{\tilde{\mu}} d\Sigma_{\tilde{\mu}} = \int_{\Sigma_\tau} \nabla_{\tilde{\mu}} v^{\tilde{\mu}} m^{\tilde{\lambda}} d\Sigma_{\tilde{\lambda}} + \int_{\partial \Sigma_{\tau}} v^{\tilde{\mu}} m^{\tilde{\lambda}} d S_{\tilde{\mu} \tilde{\lambda}},
\end{equation}
valid for any foliation $\Sigma_\tau$ and any vector field
$v^{\tilde{\mu}}$. Here $m^{\tilde{\mu}}$ is any vector field that satisfies
$m^{\tilde{\lambda}} (d\tau)_{\tilde{\lambda}} =1$, $dS_{\tilde{\mu}
  \tilde{\lambda}} = d S_{[\tilde{\mu} \tilde{\lambda}]}$ is the surface area
element, and the second term of the right hand side should be interpreted as a
limit of integrals over the boundaries of finite regions of
$\Sigma_\tau$. Applying this identity with $v^{\tilde{\mu}}= T^{\tilde{\mu}
  \tilde{\nu}} \xi_{\tilde{\nu}}$ gives
\begin{align} \label{eq:exactForce}
  \frac{d}{d\tau} &P_\tau(\vec{\xi})  = \int_{\Sigma_\tau} \nabla_{\tilde{\mu}} T^{\tilde{\mu}
    \tilde{\nu}} \xi_{\tilde{\nu}} m^{\tilde{\lambda}} d\Sigma_{\tilde{\lambda}}
  + \frac{1}{2} \int_{ \Sigma_\tau} T^{\tilde{\mu} \tilde{\nu}}
  (\mathcal{L}_{\xi} g)_{\tilde{\mu} \tilde{\nu}} m^{\tilde{\lambda}}
  d\Sigma_{\tilde{\lambda}} - \int_{\partial \Sigma_\tau} T^{(\text{self})\tilde{\nu}
    \tilde{\mu}} \xi_{\tilde{\mu}} m^{\tilde{\lambda}}
  dS_{\tilde{\nu} \tilde{\lambda}},
\end{align}
 In the last term, we've removed the matter contribution to the stress energy
 tensor, since it has compact spatial support and so does not contribute to the
 boundary integral in the asymptotic limit. Using Eq.~(\ref{eq:seEqsMaxwell}) we
 can rewrite the first term of (\ref{eq:exactForce}) in terms of the external
 field. Specializing to Killing vector fields, for which the second term vanishes,
 gives
\begin{align} \label{eq:exactForceEM}
 \frac{d}{d\tau} P_\tau(\vec{\xi}) =& \int_{\Sigma_\tau} \left( F^{(\text{ext})
   \tilde{\mu} \tilde{\nu}} \xi_{\tilde{\mu}} j_{\tilde{\nu}} \right)
 m^{\tilde{\lambda}} d\Sigma_{\tilde{\lambda}} - \int_{\partial \Sigma_\tau}
 T^{(\text{self}) \tilde{\mu} \tilde{\nu}} \xi_{\tilde{\nu}} m^{\tilde{\lambda}}  dS_{\tilde{\mu} \tilde{\lambda}}.
\end{align}

To obtain an explicit equation of motion for the worldline,
Eq. (\ref{eq:exactForceEM}) must be supplemented by the spin supplementary
condition (\ref{eq:spinsup}) that determines the relationship between the
4-velocity $u^\mu = dz^\mu/d\tau$ of the worldline and the 4-momentum
$P_B^\mu$. To incorporate this condition we proceed as follows. First,
we write down the following identities that are valid for any choice of vector
field $P_B^{\mu}$ along the worldline
\begin{subequations} \label{eq:lawofmotionidentity}
  \begin{align}
       m_B a^\kappa =& a^\kappa \left(m_B + P_B^\mu u_\mu \right) + \mathcal{P}^\mu{}_\lambda D_\tau P_B^\lambda - \mathcal{P}^\kappa{}_\nu D_\tau \left(\mathcal{P}^\nu{}_\lambda P_B^\lambda\right), \label{eq:lawofmotionidentityacc} \\
    D_\tau m_B =& D_\tau \left(m_B + P_B^\mu u_\mu\right) - u_\mu D_\tau P_B^\mu - a_\mu P_B^\mu.\label{eq:lawofmotionidentitymdot}
  \end{align}
\end{subequations}
Here $D_\tau \equiv u^\mu \nabla_\mu$ is the covariant derivative along the worldline,
$a^{\kappa} = D_\tau u^\kappa$ is the 4-acceleration, and
\begin{equation}
  \mathcal{P}^\mu{}_\lambda = \delta^\mu{}_\lambda + u^\mu u_\lambda
\end{equation}
is the projection tensor onto the space of vectors orthogonal to the
4-velocity. The second term in each of Eqs (\ref{eq:lawofmotionidentityacc}),
(\ref{eq:lawofmotionidentitymdot}) can be obtained from (\ref{eq:exactForceEM})
with the choice $\xi_{\tilde{\mu}} = g_{\tilde{\mu}}{}^\mu$ and the replacement
$d/d\tau \rightarrow D_\tau$. For the first and third terms, we use the general
identity (\ref{eq:surfaceTrick}) specialized to
\begin{equation}
  v^{\tilde{\mu}} = \sigma^\mu T^{\tilde{\mu} \tilde{\lambda}} n_{\tilde{\lambda}},
\end{equation}
where $n_{\tilde{\lambda}} = -(d\tau)_{\tilde{\lambda}}$ is the null normal to
the future null cone $\Sigma_\tau$. Using $\nabla_{\tilde{\mu}} \sigma^\mu =
-g_{\tilde{\mu}}{}^{\mu}$, Eq. (\ref{eq:seEqsMaxwell}), and the identity for any
vector field $v^{\tilde{\mu}}$:
\begin{equation}
  \int_{\Sigma_\tau} v^{\tilde{\mu}} d\Sigma_{\tilde{\mu}} = -
  \int_{\Sigma_\tau} v^{\tilde{\mu}} n_{\tilde{\mu}} m^{\tilde{\lambda}}
  d\Sigma_{\tilde{\lambda}},
\end{equation}
we obtain an expression for the bare momentum:
\begin{align} \label{eq:exactMomentum}
  P_B^{\mu}(\tau) =& D_\tau \int_{\Sigma_\tau} \sigma^\mu T^{\tilde{\mu}
    \tilde{\lambda}} n_{\tilde{\lambda}} d\Sigma_{\tilde{\mu}} +
  \int_{\Sigma_\tau} \sigma^\mu \left[F^{(\text{ext}) \tilde{\lambda}
      \tilde{\rho}} j_{\tilde{\rho}} - T^{\tilde{\mu} \tilde{\rho}}
    \nabla_{\tilde{\mu}} n_{\tilde{\rho}} m^{\tilde{\lambda}}\right]
  d\Sigma_{\tilde{\lambda}} - \int_{\partial \Sigma_\tau} \sigma^\mu
  m^{[\tilde{\lambda}} T^{\tilde{\mu}] \tilde{\rho}} n_{\tilde{\rho}}
  dS_{\tilde{\mu} \tilde{\lambda}}.
\end{align}
Using the method of Appendix \ref{app:sescaling}, one can show that the boundary
term in (\ref{eq:exactMomentum}) vanishes when we choose $\vec{m} =
\partial/\partial\tau$ in the coordinates constructed in
\ref{sec:computationSteps}. The expression (\ref{eq:exactMomentum}) can now be
substituted into the right hand sides of Eqs.~(\ref{eq:lawofmotionidentityacc})
and (\ref{eq:lawofmotionidentitymdot}) to give explicit evolution equations for the
worldline $z^\mu(\tau)$ and bare mass $m_B(\tau)$.

In Sec. \ref{sec:finalresultsRestricted} below we will describe a limit in which
the charge, mass, and size of the body all go to zero. In this limit, the right
hand sides of Eqs. (\ref{eq:exactForceEM}) and (\ref{eq:exactMomentum}) can be
expanded in terms of electromagnetic multipole moments discussed in
Sec. \ref{sec:moments}, thereby yielding the explicit form of the equation of
motion in this limit. This calculation is carried out in
Sec. \ref{sec:Computation}. Some of our calculations will proceed directly by taking
moments of the field equations (\ref{eq:maxwell}) and (\ref{eq:seEqs}), rather
than using Eqs. (\ref{eq:exactForceEM}) and (\ref{eq:exactMomentum}).

\subsection{Equation of motion for Harte's momentum} \label{sec:Hartecheck}

We now describe an alternative non-perturbative equation of motion for the
momentum of extended charged bodies in Minkowski spacetime, due to Harte
\cite{HarteExact}. It is based on Harte's generalized momentum,
\begin{equation} \label{eq:Harteforce}
  P_{H\,\tau}(\vec{\xi}) = \int_{\Sigma_\tau} T_M^{\tilde{\mu} \tilde{\nu}} \xi_{\tilde{\mu}} d\Sigma_{\tilde{\nu}} + E_\tau(\vec{\xi}).
\end{equation}
Here the first term coincides with our bare generalized momentum
(\ref{eq:harteMomentum}), but omits the self-field contribution. The second term
$E_\tau(\vec{\xi})$ is a kind of self-field contribution, and is given by Eq.(184) of
Ref. \cite{HarteExact}. It is a double integral over spacetime that is quadratic
in the source $j^{\tilde{\mu}}$, involves a Greens function, and depends on the
source only at times $\tau^\prime$ that are within a light-crossing time of
$\tau$. Its explicit form will not be needed in what follows.

Harte's non-perturbative equation of motion is
\begin{equation} \label{eq:exactForceHarteMomentum}
  \frac{d}{d\tau} P_{H\, \tau} (\vec{\xi}) = \int_{\Sigma_\tau} d\Sigma_{\tilde{\nu}} m^{\tilde{\nu}} \left(F^{\tilde{\lambda} \tilde{\rho}} - F_S^{\tilde{\lambda \tilde{\rho}}}\right) \xi_{\tilde{\lambda}} j_{\tilde{\rho}},
\end{equation}
for Killing vectors $\vec{\xi}$, where $F_S^{\tilde{\lambda} \tilde{\rho}}$ is
the average of retarded and advanced self-fields. Harte incorporates the
spin-supplementary condition by solving explicitly for the relationship between
the 4-velocity and momentum with a choice of parameter $\tau$ which differs from
proper time. We find it more convenient to proceed instead as described above
using the general identity (\ref{eq:lawofmotionidentity}) and choosing $\tau$ to
be proper time.

We shall make use of Harte's equation (\ref{eq:exactForceHarteMomentum}) as a
partial consistency check of our results. By subtracting
Eqs.(\ref{eq:exactForce}) and (\ref{eq:exactForceHarteMomentum}), we obtain
\begin{align} \label{eq:diffconfirmation}
 \int_{\Sigma_\tau} d\Sigma_{\tilde{\nu}} m^{\tilde{\nu}} F_R^{\tilde{\lambda} \tilde{\rho}}\xi_{\tilde{\lambda}} j_{\tilde{\rho}} &+ \int_{\partial \Sigma_\tau} T^{(\text{self}) \tilde{\mu} \tilde{\nu}} \xi_{\tilde{\nu}} m^{\tilde{\lambda}} dS_{\tilde{\mu} \tilde{\lambda}}  = \left(\begin{array}{c}\text{Some total}\\\text{time derivative}\end{array}\right),
\end{align}
where $F_R^{\tilde{\lambda} \tilde{\rho}}$ is the radiative self-field, one half
the retarded field minus one half the advanced field. We compute the left hand
side of explicitly in terms of our multipole expansion and verify that it is a
total time derivative at each order in the expansion; see
Secs. \ref{sec:GHWtotalderiv} and \ref{sec:secondordertotalderiv} below.

\section{The point particle limit in the electromagnetic case} \label{sec:pointparticle}

\subsection{One parameter families of solutions: the Gralla-Harte-Wald axioms} \label{sec:axioms}

We will consider a small charged body interacting with an external
electromagnetic field. To describe the limit in which the body becomes
very small, we consider a one-parameter family of solutions of the field
equations for the body, labeled by a dimensionless parameter
$\lambda$. Following GHW, we impose the following axioms on the family of
solutions. The axioms enforce that the mass and charge of the body go to zero as
the size goes to zero.

\newtheorem{axiom}{Axiom} \newtheorem{lemma}{Lemma}

\begin{axiom} \label{axiom1}
There exists a one-parameter family of fields consisting of the Maxwell tensor
$F_{\mu \nu} (\lambda, x^\mu)$, the charge current density
$j^\mu(\lambda,x^\mu)$, and the stress-energy tensor $T_M^{\mu
  \nu}(\lambda,x^\mu)$, which satisfy the Maxwell, charge current
conservation and stress-energy conservation equations:
\begin{subequations} \label{eq:axiomFieldEqs}
\begin{align}
  \nabla^\nu F_{\mu \nu}(\lambda,x^\mu) =& 4 \pi j_\nu (\lambda, x^\mu),
  \\ \nabla_{[\mu} F_{\nu \lambda]} =& 0,\\ \nabla_\mu
  j^{\mu}(\lambda, x^\mu) =& 0,\\ \nabla_\mu T^{\mu \nu}(\lambda, x^\mu)
  =& 0,
\end{align}
\end{subequations}
where $T^{\mu \nu} \equiv T^{\mu \nu}_M + T^{\mu \nu}_F$, and $T^{\mu \nu}_F$
is given by (\ref{eq:fieldSE}). These fields are defined on the
open interval $0<\lambda<\lambda_0$, for some $\lambda_0$.
\end{axiom}

\begin{axiom} \label{axiom2}
 We assume there exist functions $\tilde{j}^\mu (\lambda,t,X^{i})$ and
 $\tilde{T}_M^{\mu \nu} (\lambda,t,X^{i})$ such that for some global Lorentz
 frame coordinates $(t,x^i)$:
\begin{subequations} \label{eq:axiom2}
\begin{align} 
j^{\mu}(\lambda,t,x^i) =& \lambda^{-2} \tilde{j}^\mu \left( \lambda, t,
\frac{x^i -
  z^i(\lambda,t)}{\lambda}\right), \label{eq:tcurrdef}\\ T_M^{\mu\nu}(\lambda, t,
x^i) =& \lambda^{-2} \tilde{T}_M^{\mu \nu} \left( \lambda, t, \frac{x^i -
  z^i(\lambda,t)}{\lambda}\right) \label{eq:tsedef},
\end{align}
\end{subequations}
where $\tilde{j}^{\mu}$ and $\tilde{T}_M^{\mu\nu}$ are jointly smooth all of in
their arguments, including at $\lambda=0$, and $z^i(\lambda,t)$ is the center-of
mass worldline defined by (\ref{eq:spinsup}).
\end{axiom}

 \begin{axiom} \label{axiom3}
All of the fields $F_{\mu \nu}$, $j^{\mu}$, and $T_{\mu \nu}^M$ are jointly
smooth in $x^\mu$ and $\lambda$ away from $\lambda=0$. There exists a worldtube
$\mathcal{W}$ of compact spatial support such that the supports of
$\tilde{j}^\mu$ and $\tilde{T}^{\mu \nu}_M$ lie inside $\mathcal{W}$ for all
$\lambda$.
 \end{axiom}

 \begin{axiom} \label{axiom4}
 The external field $F^{(\text{\emph{ext}}) \mu \nu}$ defined by
   (\ref{eq:fieldSplit}) is jointly smooth in $x^{\tilde{\mu}}$ and $\lambda$,
   including at $\lambda=0$.
 \end{axiom}

\subsection{Discussion of and motivation for the axioms}

As in GHW, the axioms \ref{axiom1}-\ref{axiom4} are intended to describe a
family of physically reasonable charge current and stress-energy distributions,
such that the limit $\lambda\rightarrow0$ represents a pointlike object. At any
finite $\lambda$, however, the object is nonsingular with smooth (in particular,
non-distributional) sources and a finite self field. Our goal is to derive a set
of ordinary differential equations that govern the motion of the object in the
limit of small $\lambda$.

The axioms enforce a limit where the size $\mathcal{L}$ of the body is much
smaller than the scale\footnote{This scale can either
  be the characteristic length over which $F^{(\text{ext})}$ varies, or the
  characteristic time.} $\mathcal{L}_{\text{ext}}$ of variation of the external field $F^{(\text{ext}) \mu
  \nu}$. Thus, there is a separation of scales
\begin{equation}
 \mathcal{L} << \mathcal{L}_{\text{ext}}.
\end{equation}
One can think of the parameter $\lambda$ in our one parameter family of
solutions as being the ratio $\mathcal{L}/\mathcal{L}_{\text{ext}}$, since the
size of the body decreases linearly with $\lambda$, from
Eqs. (\ref{eq:tcurrdef}) and (\ref{eq:tsedef}). As discussed by GHW, a crucial
feature of the assumed one-parameter family is that the mass and charge of the
body go to zero as $\lambda\rightarrow 0$, at the same rate as the size.

Our axioms are identical to those of GHW except for the status of the
worldline. GHW assume the existence of a $\lambda$-independent worldline
$z^i(t)$ for which a version of (\ref{eq:axiom2}), with $z^i(\lambda, t)$
replaced by $z^i(t)$, is satisfied. By contrast, we define a one-parameter family
of worldlines $z^i(\lambda,t)$ according to the general prescription described
in Sec. \ref{sec:DHparams}. The two approaches coincide at leading order, but at
subleading order the $\lambda$-dependent worldline is more convenient.

Axiom \ref{axiom2} appears to violate Lorentz invariance by the choice of a
specific Lorentz frame. However, if this assumption is satisfied in some Lorentz
frame, it is satisfied in all Lorentz frames, so it does not violate Lorentz
invariance.

\subsection{Consequence of axioms: the near zone and far zone limits}

Following GHW, it is instructive to consider two different limits of
$\lambda\rightarrow0$ that give complementary descriptions of the interaction of
the body with the external field.

The limit $\lambda\rightarrow 0$ at fixed rescaled coordinates
\begin{equation}
  (T,X^i) \equiv \left(t, \frac{x^i - z^i(t,\lambda)}{\lambda}\right),
\end{equation}
describes the ``near zone'' limit. It describes what would be measured by
observers at distances from the object of order the object's size
$\mathcal{L}$. In this limit, points with fixed global Lorentzian coordinates
$x^i$ become more and more distant as $\lambda\rightarrow 0$. The lengthscale
$\mathcal{L}_{\text{ext}}$ of the external field goes to infinity, while the
size $\mathcal{L}$ of the body remains finite.

The limit $\lambda \rightarrow 0$ at fixed $(t,x^i)$ describes the ``far zone''
limit. It describes what would be measured by observers at distances from the
object of order $\mathcal{L}_{\text{ext}}$. In this limit, points at fixed
rescaled coordinates $(T,X^i)$ approach the worldline $x^i=z^i(0,t)$ as
$\lambda\rightarrow 0$. In particular, the object's size $\mathcal{L}\rightarrow
0$ as $\lambda\rightarrow 0$ at fixed $(t,x^i)$.

The GHW axiom approach is closely related to the matched asymptotics method
often used in gravitational calculations
\cite{Poisson,mino,pound2010,death1975,gralla2008}. The `near zone' expressions
are analogous to an expansion in positive powers of the radial coordinate, valid
near the body, and the `far zone' expressions are analogous to the expansions
approximating the body as a pointlike source.

We now discuss the limiting behavior of the self-field as $\lambda\rightarrow
0$. The assumptions of subsection \ref{sec:GovEq} do not demand smoothness of
the matter fields $j^\mu$ and $T^{\mu \nu}$ in $\lambda$ at $\lambda=0$. As
shown by GHW, it follows from axioms \ref{axiom1}-\ref{axiom4} that the limits
$\lambda \rightarrow 0$ of the matter fields $j^\mu$ and $T^{\mu \nu}$ exist as
distributions. This result reflects the desired ``point particle'' nature of the
$\lambda \rightarrow 0$ limit of the body.  However, axiom \ref{axiom4} demands
that in the limit $\lambda\rightarrow0$, the external field remains smooth in
the coordinates $x^i$. This ensures that the external field possesses a
well-defined value at the worldline, even in the point particle limit.

The limiting behavior of the self field is derived in the appendix of
\cite{GHW}, and can be described as follows. There exists a function
$\tilde{F}^{(\text{self}) \mu \nu}$, which is jointly smooth in its arguments,
including at $\lambda=0$, such that
\begin{equation}\label{eq:scaledSelfField}
F^{(\text{self}) \mu \nu}(\lambda, t, x^i) = \lambda^{-1} \tilde{F}^{(\text{self})
  \mu \nu}(\lambda,t,X^i).
\end{equation}
We define a tilded version of the full electromagnetic field $F^{\mu
  \nu}(\lambda,t,x^i)$, by
\begin{equation} \label{eq:tildedVersionFext}
\tilde{F}^{\mu \nu}(\lambda, t, X^i) = \lambda F^{\mu
  \nu}\left[\lambda,t,z^i(t,\lambda) + \lambda X^i\right].
\end{equation}
It follows from (\ref{eq:scaledSelfField}) that this full field can be written
as
\begin{align}
  \tilde{F}^{\mu \nu}(\lambda,t,X^i) =& \tilde{F}^{(\text{self}) \mu \nu}(\lambda, t, X^i) + \lambda F^{(\text{ext}) \mu \nu}(\lambda, t, z^i + \lambda X^i),
\end{align}
so as $\lambda\rightarrow0$ at fixed $X^i$, $\tilde{F}^{\mu \nu}\rightarrow
\tilde{F}^{(\text{self})\, \mu \nu}$. It also follows for
(\ref{eq:scaledSelfField}) and (\ref{eq:fieldSE}) that the stress-energy tensor
(\ref{eq:bodySE}) obeys an axiom of the form (\ref{eq:tsedef})
\begin{equation} \label{eq:rescaledSE}
  T^{\mu \nu}(\lambda, t, x^i) = \lambda^{-2} \tilde{T}^{\mu
    \nu}\left(\lambda,t,\frac{x^i - z^i(\lambda,t)}{\lambda}\right),
\end{equation}
where the right hand side is a smooth function of its arguments.

\subsection{Limiting behavior of body parameters} \label{sec:limitingbehavior}

We next specialize the general definitions (\ref{eq:baremult}) of
electromagnetic multipole moments to the one-parameter family of charge
currents. We find from Eq.(\ref{eq:tsedef}) that
\begin{subequations}\label{eq:scaledMoments}
\begin{align}
  q_B(\lambda) =&
  \lambda\tilde{q}(\lambda), \label{eq:scaledCharge}\\ \mathcal{J}_B^\mu(\tau,\lambda)
  =& \lambda
  \tilde{\mathcal{J}}^\mu(\tau,\lambda), \label{eq:scaledChargeMoment}\\ Q_B^{\mu
    \nu}(\tau,\lambda) =& \lambda^2 \tilde{Q}^{\mu
    \nu}(\tau,\lambda), \label{eq:scaledDipole}\\ Q_B^{\mu \nu
    \lambda}(\tau,\lambda) =& \lambda^3 \tilde{Q}^{\mu
    \nu}(\tau,\lambda) \label{eq:scaledQuadrupole},
\end{align}
\end{subequations}
where the rescaled moments $\tilde{q},\tilde{\mathcal{J}}^\mu,\tilde{Q}^{\mu
  \nu},$ and $\tilde{Q}^{\mu \nu \lambda}$ have Taylor expansions about
$\lambda=0$ that start at $\mathcal{O}(\lambda^{0})$, for example
\begin{equation}
  \tilde{q}(\lambda) = \tilde{q}^{(0)} + \lambda \tilde{q}^{(1)} + \dots.
\end{equation}
The result (\ref{eq:scaledMoments}) is one of the principal benefits of using
the one-parameter family of solutions: in the limit $\lambda\rightarrow 0$,
successively higher multipoles are suppressed by a higher and higher power of
$\lambda$. Hence, the limit enforces a multipole expansion.

Similar results apply to the 4-momentum $P_B^\mu$ (\ref{eq:momFlatDef}) and spin
$S_B^{\mu \nu}$ (\ref{eq:spinFlatDef}), which can be written as
\begin{subequations}
  \begin{align}
    P_B^{\mu}(\tau,\lambda) =& \lambda \tilde{P}^{\mu}(\tau,\lambda),\label{eq:scaledPdef} \\
    S_B^{\mu \nu}(\tau,\lambda) =& \lambda^2 \tilde{S}^{\mu \nu}(\tau,\lambda), \label{eq:scaledSdef}
  \end{align}
\end{subequations}
where $\tilde{P}^\mu$ and $\tilde{S}^{\mu \nu}$ have nonzero limits as $\lambda\rightarrow0$. We define a rescaled mass in terms of the rescaled momentum $\tilde{P}^\mu$,
\begin{equation} \label{eq:scaledmdef}
  \tilde{m}^2 = - \tilde{P}_{\mu} \tilde{P}^\mu,
\end{equation}
which satisfies $\lambda \tilde{m} = m_B$, and has a finite, non-zero  value in the limit $\lambda \rightarrow 0$.

\subsection{Axioms in the scalar case} \label{sec:scalarscaled}

We use a set of assumptions closely related to axioms \ref{axiom1}-\ref{axiom4} for the
scalar self force derivation. We replace the charge current $j^\mu$ with the
charge density $\rho$, the field strength $F_{\mu \nu}$ with the first
derivative of the scalar field $\Phi_{;\mu}$, and Maxwell's equations (\ref{eq:maxwell})
with the Klein-Gordon wave equation (\ref{eq:scalarfield}).

The scalar charge moments (\ref{eq:baremultScalar}) can be written as
\begin{subequations}\label{eq:scaledMomentsScalar}
\begin{align}
  q_{S B}(\lambda) =& \lambda\tilde{q}_S(\lambda), \label{eq:scaledScalarCharge}\\
  Q_{S B}^{\mu}(\tau,\lambda) =& \lambda^2 \tilde{Q}_S^{\mu
  }(\tau,\lambda), \label{eq:scaledScalarDipole}\\
  Q_{S B}^{\mu \nu}(\tau,\lambda) =& \lambda^3 \tilde{Q}_S^{\mu
    \nu}(\tau,\lambda) \label{eq:scaledScalarQuadrupole},
\end{align}
\end{subequations}
where $\tilde{q}_S$, $\tilde{Q}_S^{\mu}$, and $\tilde{Q}_S^{\mu \nu}$ have finite, non-zero limits as $\lambda\rightarrow 0$, just as for the electromagnetic moments above.

\subsection{Renormalized projected body parameters} \label{sec:restricted}

In this section we define a set of renormalized and projected body parameters -
momentum, angular momentum and electromagnetic moments - that have a number of
desirable properties:
\begin{itemize}
  \item The final equation of motion is simpler when expressed in terms of these
    body parameters rather than the original (bare) body parameters.
  \item The projected parameters have the conventional number of independent
    degrees of freedom (6 for electromagnetic dipole, 14 for quadrupole), unlike
    our original definitions (\ref{eq:baremult}) which had 10 degrees of
    freedom for the dipole and 22 for the quadrupole.
  \item The renormalizations are chosen such that the final equations of motion
    depend only on the renormalized projected parameters.
\end{itemize}

Our definitions of renormalized projected body parameters are perturbative and
are limited to the context of the one-parameter family of solutions. It would be
interesting to find more general, non-perturbative definitions that reduce to
these definitions in the $\lambda\rightarrow 0$ limit. We have been unable to do
so. We do note that our renormalized projected parameters are not all obtained
at second order by taking the $\lambda \rightarrow 0 $ limit of Harte's
non-perturbative definitions specialized to a spacelike foliation
$\Sigma_\tau$. We therefore expect that such a procedure will not hold in
general, and merely define the renormalized, projected moments perturbatively.

\def\arraystretch{2.0}
\begin{table*}
  \resizebox{\textwidth}{!}{%
  \begin{tabular}{|r|c |c| c | c | c | c | c|}
    \hline
    Bare moments& $P^\mu_B$~(\ref{eq:momFlatDef}) : 4 & $m_B$~(\ref{eq:restmass})~:~1& $S^{\mu \nu}_B$~(\ref{eq:spinFlatDef})~:~3 & $q_B$~(\ref{eq:bareCharge})~:~1& $\mathcal{J}_B^\mu$~(\ref{eq:bareChargeMoment})~:~0 & $Q_B^{\mu \nu}$~(\ref{eq:bareDipole})~:~10 & $Q_B^{\mu \nu \lambda}$~(\ref{eq:bareQuadrupole})~:~22  \\\hline
    Rescaled bare moments & $\tilde{P}$~(\ref{eq:scaledPdef}) : 4& $\tilde{m}$~(\ref{eq:scaledmdef})~:~1 & $\tilde{S}^{\mu \nu}$~(\ref{eq:scaledSdef}) : 3 & $\tilde{q}$~(\ref{eq:scaledCharge})~:~1&$\tilde{\mathcal{J}}^\mu$~(\ref{eq:scaledChargeMoment})~:~0 & $\tilde{Q}^{\mu \nu}$~(\ref{eq:scaledDipole})~:~10 & $\tilde{Q}^{\mu \nu \lambda}$~(\ref{eq:scaledQuadrupole})~:~22\\\hline
    Renormalized projected moments & Not required & $m$~(\ref{eq:restrictedMass})~:~1 &  $S^{\mu \nu}$~(\ref{eq:restrictedSpin})~:~3& q~(\ref{eq:restrictedCharge})~:~1& not required & $Q^{\mu \nu}$~(\ref{eq:restrictedDipole})~:~6&$Q^{\mu \nu \lambda}$~(\ref{eq:restrictedQuadrupole})~:~14\\\hline
  \end{tabular}}
  \caption{\label{table:momentsTable} A summary of the various body parameters we have defined. Each cell lists the symbol for the quantity, the number of the equation in which the quantity is defined, and the number of independent components in the quantity after the charge conservation and the spin supplementary condition have been imposed.}
\end{table*}
\def\arraystretch{1}

The renormalized mass is given by
\begin{align}
    m =& -\tilde{P}^\mu u_\mu - \lambda u_\mu F^{(\text{ext}) \mu}{}_{\nu} \tilde{Q}^{\nu\lambda}u_\lambda -  \tfrac{2}{3} \lambda^2 \tilde{q} a_{\mu} D_\tau \left(\mathcal{P}^\mu{}_\nu  \tilde{Q}^{\lambda \nu}u_\lambda\right)  \notag\\&- \lambda^2 u_\mu F^{(\text{ext})\mu}{}_{\nu ;\lambda} \mathcal{P}^\lambda{}_\eta  \tilde{Q}^{\nu\eta\sigma} u_\sigma   + \lambda^2 u_\mu F^{(\text{ext})\mu}{}_{\nu}  \tilde{Q}^{\nu\lambda \eta} a_{\lambda} u_\eta   +  \mathcal{O}(\lambda^3),\label{eq:restrictedMass}
\end{align}
where $u^\mu$ is the 4-velocity and $a^\mu$ the 4-acceleration of the
worldline, $\mathcal{P}_\mu{}^\nu = \delta_\mu{}^\nu + u_\mu u^\nu$ is the
projection tensor, and $D_\tau = u^\mu \nabla_\mu$. The rescaled electromagnetic
dipole $\tilde{Q}^{\mu \lambda}$ and quadrupole $\tilde{Q}^{\mu \nu \lambda}$
which appear here are defined in Eq. (\ref{eq:scaledMoments}).

Note that $\tilde{P}^{\mu} u_\mu = - \tilde{m} + \mathcal{O}(\lambda)$, so $m$
and $\tilde{m}$ coincide to leading order. In the limit $\lambda\rightarrow 0$
the renormalized mass can be expanded as
\begin{equation}
  m(\lambda) = m^{(0)} + \lambda m^{(1)} + \lambda^2 m^{(2)} + \dots,
\end{equation}
where the coefficients $m^{(0)}$,$m^{(1)}$, etc are independent of $\lambda$ and
$m^{(0)} \ne 0$.

We do not define a renormalized momentum since the momentum is eliminated in the
final equation of motion.

The renormalized spin is
\begin{align}
    S^{\mu \nu}  =& \tilde{S}^{\mu \nu} + 2 \lambda F^{(\text{ext})[\mu|}{}_\lambda \tilde{Q}^{\lambda | \nu ] \rho} u_\rho  + \tfrac{2}{3} \lambda \tilde{q} \mathcal{P}^{[\mu}{}_\lambda u^{\nu]} \tilde{Q}^{\lambda \rho} a_\rho \notag\\
  & + \lambda \mathcal{P}^{[\mu}{}_\lambda \mathcal{P}^{\nu]}{}_\rho \left(\tfrac{2}{3} \tilde{q} D_\tau \tilde{Q}^{\lambda \rho} + \tfrac{4}{3} \tilde{q} u_\eta \tilde{Q}^{\eta \lambda} a^\rho + \tfrac{2}{3} \tilde{q} \tilde{Q}^{\lambda \eta} u_\eta  a^\rho\right) +  \mathcal{O}(\lambda^2). \label{eq:restrictedSpin}
\end{align}
This also can be expanded in powers of $\lambda$ with a leading term which is non-zero.

The charge is conserved so requires no renormalization,
\begin{equation} \label{eq:restrictedCharge}
  q = \tilde{q}.
\end{equation}

The renormalized, projected electromagnetic dipole is
\begin{align}\label{eq:restrictedDipole}
  Q^{\mu \nu} =& \left(\tilde{Q}^{\mu \nu} + \lambda u_\sigma D_\tau\left( \tilde{Q}^{\mu \nu \sigma}\right)\right) \mathcal{P}_\nu{}^\kappa + \mathcal{O}(\lambda^2),
\end{align}
Note that this dipole is orthogonal to the 4-velocity on its second index, unlike the bare dipole. We can expand $Q^{\mu \nu}$ as
\begin{equation}\label{eq:expandthedipole}
  Q^{\mu \nu} = Q^{(0) \mu \nu} + \lambda Q^{(1) \mu \nu} + \mathcal{O}(\lambda^2).
\end{equation}
Charge conservation [Eq. (\ref{eq:currentIntegration}) below with $m=2$ and $N=2$]
enforces that the spatial components of the leading order term are antisymmetric,
\begin{equation}
  Q^{(0) \mu \nu} \mathcal{P}_\mu{}^{(\lambda} \mathcal{P}_\nu{}^{\eta)} = 0.
\end{equation}
At higher order, the quantity $Q^{(1) \mu \nu} \mathcal{P}_\mu{}^{(\lambda}
\mathcal{P}_\nu{}^{\lambda)}$ can be computed from the time derivative of the
electric quadrupole and the corresponding subleading charge conservation
[Eq. (\ref{eq:currentIntegration}), order $\mathcal{O}(\lambda$), with $m=2$ and $N=2$]. Hence, the dipole
(\ref{eq:restrictedDipole}) has 6 independent components.

We note that if we replace the future null cone $\Sigma_\tau$ in the definitions
(\ref{eq:baremult}) of electromagnetic moments with a spacelike hypersurface
orthogonal to the 4-velocity, then the same final result would be obtained by
taking the expression (\ref{eq:restrictedDipole}) but omitting the correction
term.

The renormalized, projected quadrupole is
\begin{align} \label{eq:restrictedQuadrupole}
  Q^{\mu \lambda \eta}  =& \mathcal{P}^\lambda{}_\nu \mathcal{P}^\eta{}_\sigma \tilde{Q}^{\mu \nu \sigma} + \mathcal{O}(\lambda).
\end{align}
This tensor is orthogonal to the 4-velocity in its second two indices. The completely symmetric part of the spatial projection of this quadrupole vanishes to leading order
\begin{equation}
 \label{eq:RestrictedQuadSymmetry}
  Q^{\mu \nu \sigma}\mathcal{P}_\mu{}^{(\lambda} \mathcal{P}_\nu{}^\eta \mathcal{P}_\sigma{}^{\rho)} =  \mathcal{O}(\lambda),
\end{equation}
from Eq.(\ref{eq:currentIntegration}) below with $m=3$, $N=3$. It follows that
the leading order renormalized quadrupole has the standard number of independent
components (6 electric and 8 magnetic).

The notations for and properties of the various body parameters we have defined
are summarized in Table \ref{table:momentsTable}.

\section{Summary of Results: Electromagnetic laws of motion through second order} \label{sec:laws}

\subsection{Preamble: domain of validity of self force equations} \label{sec:preamble}

The classic Abraham-Lorentz-Dirac radiation-reaction equation,
\begin{equation} \label{eq:ALD}
a^{\nu} = \frac{q}{m} F^{\nu \mu} u_\mu + \frac{2}{3} \frac{q^2}{m}
\mathcal{P}^\nu{}_\mu \dot{a}^{\mu},
\end{equation}
is a third-order differential equation which possesses transparently nonphysical
runaway solutions.

As pointed out by GHW, (\ref{eq:ALD}) is valid only in the regime $q^2 \dot{a}
/m a \equiv \epsilon \ll 1$, and the equation has errors of order $\epsilon^2
a$. The runaway solutions possess a rapidly growing acceleration, and violate
the assumption $\epsilon \ll 1$. When $\epsilon \gtrsim 1$, the perturbative
differential equation (\ref{eq:ALD}) is no longer a good approximation.

The reduction of order procedure provides a method of deriving from
Eq. (\ref{eq:ALD}) an equation which is equally accurate but which is second
order in time and which does not have runaway solutions
\cite{eliezer,landau,simon,flanaganwald,rohrlich}. Substituting the expression for the acceleration given by the first term in (\ref{eq:ALD}) into the second term modifies the equation by a term which is no larger than the pre-existing error terms. The resulting reduced-order equation is
\begin{align}
  a^{\sigma} =& \frac{q}{m} F^{\sigma \mu} u_\mu + \frac{2}{3}\frac{q^3}{m^2}
  \mathcal{P}^\sigma{}_\rho \left( F^{\rho \mu}{}_{;\nu} u_\mu u^\nu +
  F^{\rho \mu} F_{\mu \nu} u^\nu \right) + \mathcal{O}(q^5).
\end{align}
Our final results (\ref{eq:selfForceSeries}) are expressed as an expansion in
powers of $\lambda$, a parameter which is proportional to the charge $q$, also the
mass $m$, and here also to $q^2/m$. We do not perform a reduction of order in our
results for brevity. (except the point particle case discussed in
Sec. \ref{sec:empointparticle} below). However, we emphasize that our results
should be interpreted in terms of their reduced-order counterparts.

\subsection{Laws of motion - general self force and center of mass evolution} \label{sec:finalresultsRestricted}

We present in this section the results for the electromagnetic case. The scalar
results are derived in much the same way, and can be found in the appendix
\ref{app:ScalarLaws}.

The evolution of the body's worldline $z^{\mu}(\tau)$ and rest mass to second
order in $\lambda$ are given by
\begin{subequations} \label{eq:selfForceSeries}
  \begin{align}
    m a^{\mu} =& f^{(0)\mu} + \lambda f^{(1)\mu} + \lambda^2 f^{(2)\mu} + \mathcal{O}(\lambda^3),\\
    D_\tau m =& \lambda \mathcal{F}^{(1)} + \lambda^2 \mathcal{F}^{(2)} + \mathcal{O}(\lambda^3),
  \end{align}  
\end{subequations}
where $a^\mu$ is the acceleration of the worldline and $m$ is the renormalized
mass (\ref{eq:restrictedMass}). Here $f^{(0)\mu}$ is the Lorentz force,
$f^{(1)\mu}$ and $\mathcal{F}^{(1)}$ are the first order GHW results, and
$f^{(2)}$ and $\mathcal{F}^{(2)}$ are the new second-order results presented
here. Explicit expressions for all these quantities are given in this section
and the derivations are given in Sec.\ref{sec:Computation} below

We refer to Eqs. (\ref{eq:selfForceSeries}) as `laws' of motion, instead of
equations of motion, as they require additional information about the body's
electromagnetic multipoles their time dependence to fully determine the motion.
The requisite additional equations parameterize the evolution of the internal
degrees of freedom of the body.

At leading order we have the Lorentz force and mass conservation
\begin{subequations}
  \begin{align}
f^{(0) \mu} =& q F^{(\text{ext})\mu \lambda} u_\lambda,\\ D_\tau m =& \mathcal{O}(\lambda).
  \end{align}
\end{subequations}

At subleading order we have,
\begin{subequations} \label{eq:GHWreproduce}
  \begin{align}
    f^{(1) \kappa}  =&  \mathcal{P}^{\kappa}{}_\nu \bigg[F^{(\text{ext})\nu}{}_{\mu ;\lambda} Q{}^{ \mu \lambda}  + \tfrac{2}{3} q   D_{\tau}a^{\nu} +   D_\tau \left( a_{\mu} S^{\nu \mu} \right) \notag\\& \hspace{1cm} + F^{(\text{ext})\nu}{}_\mu D_\tau Q^{[\mu \lambda]} u_\lambda -   D_\tau \left(u_\mu F^{(\text{ext})\mu}{}_\lambda Q^{\lambda \nu}  \right)  \bigg], \\
    \mathcal{F}^{(1)}=& -u_\mu F^{(\text{ext}) \mu}{}_{\nu;\lambda} Q{}^{\nu \lambda} - u_\nu  F^{(\text{ext})\nu}{}_\mu D_\tau \left(Q{}^{\mu \lambda}  \right) u_\lambda - 2 u_\mu F^{(\text{ext}) \mu}{}_\nu Q^{\nu \lambda} a_\lambda. 
  \end{align}
\end{subequations}
Here the body's charge $q$, electromagnetic dipole $Q^{\mu \nu}$, and spin $S^{\mu \nu}$ are the renormalized versions (\ref{eq:restrictedCharge}), (\ref{eq:restrictedDipole}), and (\ref{eq:restrictedSpin}).

To facilitate comparison of the results with those of GHW, we define an
antisymmetric dipole $Q_A^{\mu \nu}$ by
\begin{subequations} \label{eq:antisymmetricDipole}
  \begin{align}
    Q_A{}^{\mu \nu} \mathcal{P}_\nu{}^\lambda =& Q^{\mu \lambda},\\
    Q_A{}^{\mu \nu} u_\nu =& - u_\nu Q^{\nu \mu},
  \end{align}
\end{subequations}
for which $Q_A{}^{(\mu \nu)} = 0$. Eliminating  $Q^{\mu \nu}$ in terms of $Q_A^{\mu \nu}$, and we find
\begin{subequations} 
  \begin{align}
    f^{(1) \kappa} =&  \mathcal{P}^{\kappa}{}_\nu \bigg[F^{(\text{ext})\nu}{}_{\mu ;\lambda} Q_A{}^{ \mu \lambda}  + \tfrac{2}{3} q   D_{\tau}a^{\nu} +   D_\tau \left( a_{\mu} S^{\nu \mu} \right)    +  2 D_\tau \left(u_\mu F^{(\text{ext})\lambda [\nu} Q_A{}^{\mu]}{}_\lambda  \right) \bigg], \label{eq:GHWselfForce} \\
    \mathcal{F}^{(1)}=& -u_\mu F^{(\text{ext}) \mu}{}_{\nu;\lambda} Q_A{}^{\nu \lambda} - D_\tau \left(F^{(\text{ext})\nu}{}_\mu Q_A{}^{\mu \lambda}  \right)u_\nu u_\lambda  - 2 u_\mu F^{(\text{ext}) \mu}{}_\nu Q_A{}^{\nu \lambda} a_\lambda, \label{eq:GHWmassev} 
  \end{align}
\end{subequations}
which agrees with the results of GHW.  The third term in the mass evolution
(\ref{eq:GHWmassev}) does not appear in GHW, however it gives only a
$\mathcal{O}(\lambda^2)$ contribution when reduction of order is applied. We
retain this term since we will be working to $\mathcal{O}(\lambda^2)$.

As noted in GHW, the first and second terms in the acceleration equation
(\ref{eq:GHWselfForce}) are the monopole self force usually derived from the
radiative self field, and the direct interactions with the external field. The
final two  terms in (\ref{eq:GHWselfForce}) are terms that are not usually
derived in elementary treatments of electrodynamics.

The second order results can be decomposed into monopole, dipole, and quadrupole contributions:
\begin{subequations} \label{eq:secondorderstart}
  \begin{align}
    f^{(2)\mu} =& f^{(2)\mu}_{\text{point}} + f^{(2)\mu}_{\text{dipole}} + f^{(2)\mu}_{\text{quadrupole}},\\
    \mathcal{F}^{(2)} =& \mathcal{F}^{(2)}_{\text{point}} + \mathcal{F}^{(2)}_{\text{dipole}} + \mathcal{F}^{(2)}_{\text{quadrupole}}.
  \end{align}
\end{subequations}
We have
\begin{subequations}
  \begin{align}
    f^{(2) \mu}_{\text{point}} = 0,\\
    \mathcal{F}^{(2)}_{\text{point}} = 0,
  \end{align}
\end{subequations}
so there are no new point particle terms at second order. We note, however, that
monopole terms at $\mathcal{O}(\lambda^2)$ would be generated if one expands out
the body parameters in a power series in $\lambda$, as in
Eq. (\ref{eq:expandthedipole}) above, and also would be generated by the
reduction of order procedure, c.f. Sec.\ref{sec:empointparticle} below.  The
explicit, new, dipole and quadrupole contribution to the self force are
\begin{subequations} 
  \begin{align}
   f^{(2) \mu}_{\text{dipole}} = \mathcal{P}^\sigma{}_\kappa\bigg[&  - \tfrac{1}{3} q a_{\mu} a^{\mu} a_{\nu} Q^{\nu \kappa} + q a^{\kappa} D_{\tau}a^{\mu} \mathcal{P}_{\mu \nu} Q^{\lambda \nu} u_\lambda  +  \tfrac{7}{6} q D_{\tau}a^{\kappa} a_{\mu}  Q^{\lambda \mu} u_\lambda   \nonumber \\
  & - \tfrac{11}{6} q a_{\mu} D_{\tau}a^{\mu}  Q^{\nu\kappa} u_\nu  +  \tfrac{1}{3}q a^\kappa a_{\mu} D_{\tau}Q^{\nu \mu}u_\nu -  q a_{\mu} a^{\mu}  D_{\tau}Q^{\nu \kappa} u_\nu  \nonumber \\
  &   -  \tfrac{2}{3} q D_{\tau}a_{\mu}  D_{\tau}Q^{\mu \kappa}  -  2 q a_{\mu} D_{\tau}{}^2Q^{\mu \kappa}  - \tfrac{2}{3} q D_{\tau}{}^3Q^{\mu \kappa} u_\mu  \bigg]
,\\
      f^{(2) \mu}_{\text{quadrupole}}  = 
 \mathcal{P}^\sigma{}_\kappa\bigg[&  \tfrac{1}{2} F^{(\text{ext})\kappa}{}_{\mu;\nu \lambda} Q{}^{\mu \nu \lambda}    -  u_\mu D_\tau \left( F^{(\text{ext}) \mu}{}_{\nu;\rho} \mathcal{P}^\nu{}_\lambda Q{}^{\lambda\kappa\rho}  \right)     + \tfrac{1}{2} D_{\tau}{}^2 \left(F^{(\text{ext})\kappa}{}_\mu Q{}^{\mu \rho}{}_\rho\right) \notag\\&- 2  u_\mu F^{(\text{ext})\mu}{}_{\lambda;\nu} u^\nu Q{}^{\lambda\kappa\rho} a_{\rho}   + 2  F^{(\text{ext})[\kappa}{}_{\mu;\lambda} Q{}^{\mu|\nu]\lambda} a_{\nu} +  \tfrac{1}{2} F^{(\text{ext})\kappa}{}_{\mu;\nu} a^{\nu} Q{}^{\mu\rho}{}_\rho \notag\\& - \tfrac{1}{2} F^{(\text{ext})\kappa \nu}  D_{\tau}\left(a_{\nu} u_\mu Q{}^{\mu \rho}{}_{\rho}\right)  -  u_\mu F^{(\text{ext})\mu}{}_\nu  D_{\tau}\left(Q{}^{\nu \kappa \lambda}  a_{\lambda}\right) + a^{\kappa} u_\mu F^{(\text{ext})\mu}{}_{\nu} D_{\tau}Q{}^{\nu \rho}{}_\rho \nonumber \\
 &  - 2  a_{\nu}  F^{(\text{ext})(\nu}{}_{\mu} Q{}^{\mu |\kappa)\lambda} a_{\lambda}  \bigg],
  \end{align}
\end{subequations}
and the explicit, new, dipole and quadrupole contributions to the mass evolution are
\begin{subequations} \label{eq:diquadMdotSecond}
  \begin{align}
    \mathcal{F}^{(2)}_{\text{dipole}} =& -\tfrac{1}{3}\tilde{q} a_{\mu} a^{\mu}  u_\lambda Q_{R}{}^{\lambda \nu} a_{\nu}  - \tfrac{2}{3} \tilde{q} D_{\tau}a_{\nu} \mathcal{P}^\nu{}_\lambda  D_{\tau}\left( u_\mu Q{}^{\mu  \lambda} \right) , \\
     \mathcal{F}^{(2)}_{\text{quadrupole}}  =& - \tfrac{1}{2} u_\mu F^{(\text{ext})\mu}{}_{\lambda;\nu \rho} Q{}^{\lambda \nu \rho} -  \tfrac{1}{2} u_\mu   F^{(\text{ext})\mu}{}_{\lambda;\nu \sigma} u^\nu u^\sigma Q{}^{\lambda\rho}{}_\rho - 2 u_\mu F^{(\text{ext})\mu}{}_{\lambda;\sigma}  Q{}^{\lambda \sigma\nu} a_{\nu} \nonumber \\
  &  - u_\mu F^{(\text{ext})\mu}{}_{\nu;\lambda} a^{\lambda} Q{}^{\nu \rho}{}_\rho  -  \tfrac{1}{2} D_{\tau}a_{\mu} F^{(\text{ext})\mu \nu}u_\nu  u_\lambda Q{}^{\lambda\rho}{}_\rho  - \tfrac{1}{2} a_{\nu} F^{(\text{ext})\nu\mu} u_\mu a_{\lambda}  Q{}^{\lambda \rho}{}_\rho \notag\\
  & -  \tfrac{1}{2} a_{\lambda}  F^{(\text{ext})\lambda\mu} u_\mu u_\nu D_{\tau}Q{}^{\nu \rho}{}_\rho     + a_{\nu} F^{(\text{ext})\nu}{}_\mu D_{\tau}Q{}^{\mu\rho}{}_\rho   + \tfrac{1}{2} u_\mu F^{(\text{ext})\mu}{}_\lambda D_{\tau}{}^2Q{}^{\lambda \rho}{}_\rho.
  \end{align}
\end{subequations}

\subsection{Laws of motion - evolution of spin}

Like the self force, the torque may also be written in terms of the
renormalized dipole, quadrupole and spin introduced in
Sec.\ref{sec:restricted}. The result is
\begin{align} \label{eq:restrictedTorque}
  D_\tau S^{\lambda \rho} \mathcal{P}_\lambda{}^\kappa \mathcal{P}_\rho{}^\sigma =& \mathcal{P}^\kappa{}_\lambda \mathcal{P}^\sigma{}_\rho \bigg(  2  F^{(\text{ext})[\lambda}{}_{\mu} Q{}^{\mu | \rho]}  + 2 \lambda F^{(\text{ext})[\lambda}{}_{\nu; \mu} Q{}^{\nu \mu | \rho]} \notag\\& \hspace{2cm}- \tfrac{4}{3} \lambda q D_{\tau}a^{[\lambda} Q{}^{\mu| \rho]} u_\mu  +  2 \lambda F^{(\text{ext})[\lambda}{}_{\mu} Q{}^{\mu \nu |\rho]} a_{\nu}\bigg) + \mathcal{O}(\lambda^2).
\end{align}
Because of the spin supplementary condition (\ref{eq:spinsup}), this projected
version of $D_\tau S^{\lambda \rho}$ is sufficient to determine the entire time
derivative.  The first term in this torque expression reproduces the GHW result.

\subsection{Laws of motion - reduced order point particle limit} \label{sec:empointparticle}

In this section, we specialize to monopole bodies, i.e. those with vanishing
spin $S^{\mu \nu}$, electromagnetic dipole $Q^{\mu \nu}$, and electromagnetic
quadrupole $Q^{\mu \nu \lambda}$. The equations of motion
(\ref{eq:selfForceSeries}) then reduce to
\begin{subequations}
  \begin{align}
    m a^{\mu} =&  \lambda q  F^{(\text{ext}) \mu \lambda} u_\lambda +  \tfrac{2}{3} \lambda^2 q^2 \mathcal{P}^\mu{}_\nu D_{\tau}a^{\nu}  + \mathcal{O}(\lambda^4),\\
    D_\tau  m =& \mathcal{O}(\lambda^3).
  \end{align}
\end{subequations}

We now apply a reduction of order to determine the acceleration through
$\mathcal{O}(\lambda^2)$ in terms of the external field. The resulting
acceleration, given explicitly for the first time, is
\begin{align} \label{eq:ppselfForce}
      a^\mu  =& \frac{q}{m} F^{(\text{ext})\mu \nu} u_\nu + \frac{2 q^3}{3 m^2}  D_\tau F^{(\text{ext})\mu \nu} u_\nu  + \frac{2 q^4 \lambda}{3 m^3} \mathcal{P}^\mu{}_\eta F^{(\text{ext}) \eta \nu} F^{(\text{ext})}{}_{\nu \sigma} u^\sigma \notag\\
      & + \frac{4 q^5}{9 m^3} \lambda^2 D_\tau{}^2 F^{(\text{ext}) \mu \nu} u_\nu +  \frac{4 q^6}{9 m^4} \lambda^2 \mathcal{P}^\mu{}_\rho  \bigg( 2 D_\tau F^{(\text{ext})\rho \nu} F^{(\text{ext})}{}_{\nu \lambda} u^\lambda +   F^{(\text{ext})\rho \nu} D_\tau F^{(\text{ext})}{}_{\nu \lambda} u^\lambda  \bigg)\notag\\
      & +   \frac{4 q^7}{9 m^5} \lambda^2  \mathcal{P}^\mu{}_\rho  F^{(\text{ext}) \rho \nu} \mathcal{P}_{\nu \eta}  F^{(\text{ext}) \eta \lambda} F^{(\text{ext})}{}_{\lambda \sigma} u^\sigma + \mathcal{O}\left(\lambda^3\right).
\end{align}

\section{Details of derivation} \label{sec:Computation}

\subsection{Preliminary definitions and constructions}  \label{sec:computationSteps}

The derivation is based on the axioms described in sec \ref{sec:axioms}, which
are expressed in some global Lorentz frame coordinates $(t,x^i)$. For the
purposes of our derivation, we adopt a retarded body-following coordinate
system, motivated by the scaled coordinates $(T,X^i)$ considered in
Sec. \ref{sec:axioms}.

We choose a tetrad at a point on the worldline, $z^\mu(\tau,\lambda)$\footnote{Note that our construction is based on the $\lambda$-dependent worldline $z^{\mu}(\tau,\lambda)$, and not on the fixed, $\lambda$-independent worldline $z^{\mu}(\tau,0)$.},
\begin{equation}
  \{e_{\hat{0}}{}^\mu,e_{\hat{i}}{}^\mu\} \equiv \{u^\mu,e^\mu{}_{\hat{i}}\},
\end{equation}
which we constrain to be orthonormal:
\begin{equation}
  \vec{e}_{\hat{a}} \cdot \vec{e}_{\hat{b}} = \eta_{\hat{a} \hat{b}}.
\end{equation}
We extend this tetrad along the worldline using Fermi-Walker transport
\begin{equation}\label{eq:FTtetrad}
  \frac{D e^\mu{}_{\hat{a}}}{d\tau} =  e^{\nu}_{\hat{a}} \left(u^\mu a_\nu - a^\mu u_\nu\right),
\end{equation}
and extend it off the worldline by parallel transport along generators of future null cones that originate on the worldline.

Tetrad indices are raised and lowered using $\eta_{\hat{a} \hat{b}}$:
\begin{equation}
  u^\mu = e^{\mu}{}_{\hat{0}} = - e^{\mu \hat{0}} \hspace{2cm} e^{\mu}{}_{\hat{j}} = e^{\mu \hat{i}}\delta_{\hat{i} \hat{j}}.
\end{equation}

We next define the retarded Fermi coordinate system $(T,y^{\hat{i}})$ following
Poisson \cite{Poisson}. For a given spacelike point $x^{\tilde{\mu}}$, we define
$\tau(x^{\tilde{\mu}})$ such that $z^{\mu}(\tau)$ is the intersection of the
past lightcone of $x^{\tilde{\mu}}$ with the worldline, so that
\begin{equation}
  \sigma(z^\mu(\tau(x)),x^{\tilde{\mu}}) = 0.
\end{equation}
Surfaces of constant $\tau$ are future light cones of points on the worldline. We define the spatial coordinates $y^i$ by
\begin{equation}
  y^{\hat{i}} = -\delta^{\hat{i} \hat{j}} e_{\hat{j}}{}^\mu(\tau) \sigma_\mu(z_\tau,x),
\end{equation}
evaluated at $\tau=\tau(x)$. In these coordinates the metric takes the form
\cite{Poisson}
\begin{align}
  ds^2 =& -(\varphi^2 - r^2 a^2) d\tau^2 + (\delta_{\hat{i} \hat{j}} - n_{\hat{i}} n_{\hat{j}})d y^{\hat{i}} dy^{\hat{j}} + 2 (r a_{\hat{i}} - \varphi n_{\hat{i}}) dx^{\hat{i}} d\tau,
\end{align}
where $r^2 = \delta_{\hat{i} \hat{j}} y^{\hat{i}} y^{\hat{j}}$, $\varphi=1 +
y^{\hat{i}}a_{\hat{i}}$, $n^{\hat{i}} = y^{\hat{i}}/r$. The orthonormal basis in
these coordinates is given by
\begin{subequations}
  \begin{align}
    \vec{e}_{\hat{0}} =& \partial_{\tau} - r a^{\hat{i}} \partial_{\hat{i}},\\
    \vec{e}_{\hat{i}} =& \left(\delta_{\hat{i}}{}^{\hat{j}} + r n_{\hat{i}} a^{\hat{j}}\right)\partial_{\hat{j}} - n_{\hat{i}} \partial_{\tau}.
  \end{align}
\end{subequations}

Next we re-express axiom \ref{axiom2} of Sec. \ref{sec:axioms} in terms of these
coordinates and the orthonormal basis components of the tensors. From
Eq. (\ref{eq:rescaledSE}), it takes the form
\begin{subequations} \label{eq:retardedScaledSEcurr}
  \begin{align}
    T^{\hat{a} \hat{b}}(\lambda, \tau, y^{\hat{i}}) =& \lambda^{-2} \tilde{T}^{\hat{a} \hat{b}}\left(\lambda,\tau,y^{\hat{i}}/\lambda\right),\\
    j^{\hat{a}}(\lambda,\tau,y^{\hat{i}}) =& \lambda^{-2} \tilde{j}^{\hat{a}}\left(\lambda,\tau,y^{\hat{i}}/ \lambda\right),
  \end{align}
\end{subequations}
where the right hand sides are smooth functions of their arguments [distinct
  from the functions in (\ref{eq:tcurrdef}) and (\ref{eq:rescaledSE})].

Finally, we can write the rescaled body parameters of
Sec. \ref{sec:limitingbehavior} in terms of the functions $\tilde{T}^{\hat{a}
  \hat{b}}$ and $j^{\hat{a}}$:
\begin{subequations} \label{eq:retardedmomentumspin}
  \begin{align}
    \tilde{P}^{\hat{a}} =& \int d^3 Y \left(\tilde{T}^{\hat{a} \hat{0}} - \tilde{T}^{\hat{a} \hat{i}} n_{\hat{i}}\right),\label{eq:retardedmomentum}\\
    \tilde{S}^{\hat{a} \hat{b}} =& 2 \int d^3 Y R \left(n^{[\hat{a}} \tilde{T}^{\hat{b}] \hat{0}} - n^{[\hat{a}} \tilde{T}^{\hat{b}] \hat{i}} n_{\hat{i}} \right) \label{eq:retardedspin},
  \end{align}
\end{subequations}
and
\begin{subequations} \label{eq:retardedmoments}
  \begin{align}
    \tilde{q} =& \int d^3 Y (\tilde{j}^{\hat{0}} - \tilde{j}^{\hat{i}} n_{\hat{i}}), \label{eq:retardedCharge}\\
    \tilde{\mathcal{J}}^{\hat{a}} =& \int d^3 Y \tilde{j}^{\hat{a}},\\
    \tilde{Q}^{\hat{a} \hat{b}} =& \int d^3 Y R \tilde{j}^{\hat{a}} n^{\hat{b}},\\
    \tilde{Q}^{\hat{a} \hat{b} \hat{c}} =& \int d^3 Y R^2 \tilde{j}^{\hat{a}} n^{\hat{a}} n^{\hat{b}},
  \end{align}
\end{subequations}
where $Y^{\hat{i}}=y^{\hat{i}}/\lambda$, $R^2 = \delta_{\hat{i} \hat{j}}
Y^{\hat{i}} Y^{\hat{j}}$, and $\vec{n} = \vec{u} +
n^{\hat{i}}\vec{e}_{\hat{i}}$.  Here the integrals are over surfaces of constant
$\tau$, i.e. the future light cones.

\subsection{Retarded and advanced self-field} \label{sec:retadvfield}

In this subsection, we compute the near-zone expansion of the retarded field in
terms of the scaled multipoles (\ref{eq:scaledMoments}) and the retarded
coordinates from Sec. \ref{sec:computationSteps}. The computation is used in
sections \ref{sec:firstOrderDerivation}-\ref{sec:2ndorderderive}.

Consider a field point $x^{\tilde{\mu}}$. Recall that $\tau(x^{\tilde{\mu}})$
denotes the proper time at which the past lightcone of $x^{\tilde{\mu}}$
intersects the wordline $z^\mu(\tau)$. We denote by
$\mathcal{W}_-(x^{\tilde{\mu}})$ the intersection of the interior of the past
lightcone of $x^{\tilde{\mu}}$ and the worldtube $\mathcal{W}$ of the body. The
retarded, Lorenz-gauge self-field of the body can be written as
\begin{align} \label{eq:Greensfunc}
  A^{\tilde{\mu}}_-(x) =& \int d^4 x^\prime \sqrt{-g(x^\prime)} G_-{}^{\tilde{\mu}}{}_{\nu^\prime} (x,x^\prime)
  j^{\nu^\prime}(x^\prime) \notag\\=& \int_{\mathcal{W}_-} d^4 x^\prime
  g^{\tilde{\mu}}{}_{\nu^\prime}(x,x^\prime) \delta(\sigma(x,x^\prime))
  j^{\nu^\prime}(x^\prime),
\end{align}
where $G_-^\mu{}_\nu(x,x^\prime)$ is the retarded propagator in Lorenz gauge.
Here, $g^\mu{}_{\nu^\prime}$ is the parallel propagator, and the 1-dimensional
delta function $\delta(\sigma(x,x^\prime))$ constrains the integral to the
three-surface formed by the past null cone of the field point $x$.

To relate the right hand side of (\ref{eq:Greensfunc}) to the bare multipoles
(\ref{eq:baremult}), we wish to write the integral (\ref{eq:Greensfunc}) as a
series of integrals over the future null cone of the intersection point of the
center-of-mass worldline (\ref{eq:spinsup}) and the past null cone of
$x^{\tilde{\mu}}$, which we will write as $z(\tau)$.

To this end, we write $x^{\tilde{\mu}}= (\tau,y^{\hat{i}})$ and $x^{\prime
  \tilde{\mu}^\prime} = (\tau^\prime,y^{\hat{i}})$ in the retarded coordinates
of Sec. \ref{sec:computationSteps} above. We denote the value of $\tau^\prime$
at which $\sigma$ vanishes as
\begin{equation}
  \tau^\prime = \tau + \Delta \tau(\tau,y^{\hat{i}},y^{\prime \hat{i}}).
\end{equation}
The $\delta$-function $\delta(\sigma)$ can now be written as
\begin{equation} \label{eq:deltafunction}
  \delta(\sigma(x^{\tilde{\mu}},x^\prime{}^{\tilde{\nu}^\prime})) = \frac{\delta(\tau^\prime - \tau - \Delta \tau)}{|\sigma_{,\tau}(\tau,y^{\hat{i}};\tau+\Delta \tau,y^{\prime \hat{i}})|}.
\end{equation}
Inserting this into Eq. (\ref{eq:Greensfunc}), using the fact that
$|\text{det}(g_{\alpha \beta})| = 1$ in the retarded coordinates, and
multiplying by a parallel propagator factor gives
\begin{align} \label{eq:greensexp1}
A^{\tilde{\mu}}_-(\tau,y^{\hat{i}}) g_{\tilde{\mu}}{}^\mu(\tau,0;\tau,y^{\hat{i}}) = \int d^3 y^\prime \frac{g^\mu{}_{\tilde{\nu}^\prime}(\tau,0;\tau+\Delta\tau,y^{\prime \hat{i}}) j^{\tilde{\nu}^\prime}(\tau+\Delta\tau,y^{\prime \hat{i}})}{|\sigma_{,\tau^\prime}(\tau,y^{\hat{i}};\tau+\Delta\tau,y^{\prime \hat{i}})|}.
\end{align}
We now rewrite this expression in terms of the rescaled spatial coordinates
$Y^{\hat{i}} = y^{\hat{i}}/\lambda$, $Y^{\prime \hat{i}}=y^{\prime
  \hat{i}}/\lambda$ and in terms of the tilded version of the charge current
from Eq. (\ref{eq:retardedScaledSEcurr}). Noting that $\Delta \tau(\tau,\lambda
Y^{\hat{i}},\lambda Y^{\prime \hat{i}})$ vanishes as $\lambda \rightarrow 0$ at
fixed $Y^{\hat{i}}$,$Y^{\prime \hat{i}}$, we write this quantity as
\begin{equation}
  \Delta(\tau,\lambda Y^{\hat{i}},\lambda Y^{\prime \hat{i}}) = \lambda \widetilde{\Delta
    \tau}(\tau,Y^{\hat{i}},Y^{\prime \hat{i}},\lambda),
\end{equation}
where $\widetilde{\Delta \tau}$ is finite as $\lambda \rightarrow 0$. The result is
\begin{align}
  A_-^{\tilde{\mu}}(\tau,Y^{\hat{i}}) = \lambda \int d^3 Y^\prime& \bigg[g^{\tilde{\mu}}{}_{\tilde{\nu}^\prime} (\tau,\lambda Y^{\hat{i}}; \tau+ \lambda \widetilde{\Delta \tau},\lambda Y^{\prime \hat{i}})  \frac{\tilde{j}^{\tilde{\nu}^\prime}(\tau + \lambda \widetilde{\Delta \tau},Y^{\prime \hat{i}})}{|\sigma_{,\tau^\prime}(\tau,\lambda Y^{\hat{i}}; \tau + \lambda \widetilde{\Delta \tau},\lambda Y^{\prime \hat{i}})|} \bigg] .
\end{align}
Finally, we expand the right hand side in powers of $\lambda$, and we also take
the large $R=|Y|$ limit. Expressing the result in terms of components on the
orthonormal tetrad, the retarded field can naturally be expressed in terms of
the rescaled electromagnetic moments (\ref{eq:retardedmoments})
\begin{align} 
  A_-^{\hat{a}} =& \frac{\tilde{\mathcal{J}}^{\hat{a}}}{R} + \frac{\tilde{Q}^{\hat{a}\hat{j}} n_{\hat{j}}}{R^2} + \lambda a_{\hat{i}}n^{\hat{i}} \frac{\tilde{Q}^{\hat{a} \hat{j}} n_{\hat{j}}}{R} + \lambda (a^{\hat{a}} u_{\hat{b}} - u^{\hat{a}}a_{\hat{b}})\frac{\tilde{Q}^{\hat{b} \hat{0}} - \tilde{Q}^{\hat{b} \hat{j}} n_{\hat{j}}}{R} - \lambda \frac{\tilde{Q}^{\hat{a} \hat{j}} a_{\hat{j}}}{R}\notag\\& + \lambda \frac{\partial_\tau \tilde{Q}^{\hat{a} \hat{j}} n_{\hat{j}}}{R} - \lambda \frac{\partial_\tau \tilde{Q}^{\hat{a} \hat{0}}}{R} + \mathcal{O}\left(\frac{\lambda^n}{R^m}\right), \label{eq:retardedField}
\end{align}
where the omitted terms satisfy $n+m\ge3$.

We use the result (\ref{eq:retardedField}) to evaluate certain boundary terms at
infinity that arise in Sec. \ref{sec:momentsoffieldeqs} below. 

\subsection{Moments of the field equations} \label{sec:momentsoffieldeqs}

We next express the fundamental equation (\ref{eq:seEqsMaxwell}) and charge
current conservation $\nabla_{\mu} j^\mu = 0$ in terms of the coordinates
$(\tau,Y^{\hat{i}})$, using the tilded functions on the right hand sides of
(\ref{eq:retardedScaledSEcurr}). We use tetrad component of the tensors but
write the derivatives in terms of the partial derivatives with respect to the
coordinates; this unusual combination is the most convenient for our
derivation. The result is:
\begin{subequations} \label{eq:frameSEcons}
  \begin{align}
\lambda F^{(\text{ext}) \hat{k} \hat{i}} \tilde{j}_{\hat{i}} + \lambda F^{(\text{ext}) \hat{k} \hat{0}} \tilde{j}_{\hat{0}}  =&  T^{\hat{k}\hat{j}}{}_{,\hat{j}} + \lambda T^{\hat{k}\hat{0}}{}_{,0} -  \lambda n_{\hat{i}}  T^{\hat{k}\hat{i}}{}_{,0}+   \lambda a^{\hat{k}} T^{\hat{0}\hat{0}} + \lambda a_{\hat{i}} T^{\hat{k}\hat{i}} -  \lambda a_{\hat{i}} n^{\hat{i}} T^{\hat{k}\hat{0}} \notag\\&-  \lambda a^{\hat{k}} n_{\hat{i}} T^{\hat{i}\hat{0}}    -  \lambda a^{\hat{i}} R\,  T^{\hat{k}\hat{0}}{}_{,\hat{i}}  + \lambda a^{\hat{i}} n_{\hat{j}} R\,   T^{\hat{k}\hat{j}}{}_{,\hat{i}}, \\
\lambda F^{(\text{ext}) \hat{0} \hat{i}} \tilde{j}_{\hat{i}} =&  T^{\hat{i}\hat{0}}{}_{,\hat{i}} + \lambda T^{\hat{0}\hat{0}}{}_{,0} -   \lambda \hat{n}_{\hat{i}}  T^{\hat{i}\hat{0}}{}_{,0} + 2 \lambda a_{\hat{i}} T^{\hat{i}\hat{0}} - \lambda a_{\hat{i}} \hat{n}^{\hat{i}} T^{\hat{0}\hat{0}}  \notag\\& -  \lambda a_{\hat{i}} \hat{n}_{\hat{j}} T^{\hat{i}\hat{j}}   -  \lambda a^{\hat{i}} R\,  T^{\hat{0}\hat{0}}{}_{,\hat{i}}  + \lambda a^{\hat{i}} \hat{n}_{\hat{j}} R\,  T^{\hat{j}\hat{0}}{}_{,\hat{i}},
  \end{align}
\end{subequations}
and
\begin{align} \label{eq:framecurrentcons}
  0 = \delta^{i}{}_{\hat{j}} j^{\hat{j}}{}_{,i} + \lambda j^{\hat{0}}{}_{,0} -  \lambda n^{\hat{i}} \delta_{\hat{i}\hat{j}} j^{\hat{j}}{}_{,0} +  \lambda a_{\hat{i}} j^{\hat{i}} -  \lambda a_{\hat{i}} j^{\hat{0}} n^{\hat{i}}   -  \lambda a^{\hat{i}} R\, \delta^{j}{}_{\hat{i}} j^{\hat{0}}{}_{,j} + \lambda a^{\hat{i}} n^{\hat{j}} R\, \delta_{\hat{j}\hat{k}} \delta^{l}{}_{\hat{i}} j^{\hat{k}}{}_{,l},
\end{align}
where $f_{,0}$ means $\partial f/\partial \tau$ and $\partial_{\hat{i}} f$ means
$\partial f/\partial y^{\hat{i}}$.

We next multiply (\ref{eq:frameSEcons}) and (\ref{eq:framecurrentcons}) by $R^m n^{\hat{j}_1} \dots n^{\hat{j}_N}$ for integers $m$ and $N$ and integrate with respect to $Y$. this gives the hierarchy of moment equations
\begin{subequations} \label{eq:integrationSet}
  \begin{align}
    \int d^3 Y \nabla_\mu \tilde{T}^{\hat{i} \mu} R^m n^{\hat{j}_1} \dots n^{\hat{j}_N} =\;& \int d^3 Y F^{(\text{ext})\hat{i} \mu} j_\mu R^m n^{\hat{j}_1} \dots n^{\hat{j}_N},\label{eq:spatialseIntegration}\\
    \int d^3 Y \nabla_\mu \tilde{T}^{\hat{0} \mu} R^m n^{\hat{j}_1} \dots n^{\hat{j}_N} =\;& \int d^3 Y F^{(\text{ext}) \hat{0} \mu} j_\mu R^m n^{\hat{j}_1} \dots n^{\hat{j}_N},\label{eq:timeseIntegration}\\
    \int d^3 Y \nabla_\mu \tilde{j}^{\mu} R^m n^{\hat{j}_1} \dots n^{\hat{j}_N} =\;& 0.\label{eq:currentIntegration}
  \end{align}
\end{subequations}
In these equations the arguments of all of the functions are
$(\lambda,\tau,Y^{\hat{i}})$, except for $F^{(\text{ext}) \hat{a} \hat{b}}$, for
which the arguments are as on the right hand side of
Eq. (\ref{eq:tildedVersionFext}).

We now expand the $\lambda$-dependence of $\tilde{T}^{\hat{a} \hat{b}}$ and
$\tilde{j}^{\hat{a}}$ at fixed $(\tau,Y^{\hat{i}})$ as
\begin{subequations}
  \begin{align}
    \tilde{T}^{\hat{a} \hat{b}} = \tilde{T}^{(0) \hat{a} \hat{b}} + \lambda \tilde{T}^{(1) \hat{a} \hat{b}} + \mathcal{O}(\lambda^2)\\
    \tilde{j}^{\hat{a}} = \tilde{j}^{(0) \hat{a}} + \lambda \tilde{j}^{(1) \hat{a}} + \mathcal{O}(\lambda^2),
  \end{align}
\end{subequations}
with corresponding expansion of the rescaled moments
\begin{equation}
  \tilde{P}^{\hat{a}} = \tilde{P}^{(0) \hat{a}} + \lambda \tilde{P}^{(1) \hat{a}} + \mathcal{O}(\lambda^2),
\end{equation}
and similarly for each of the spin (\ref{eq:retardedspin}) and the electromagnetic
moments (\ref{eq:retardedmoments}).

The first moments of the spatial component (\ref{eq:spatialseIntegration})
at leading order, after integrating the spatial partial derivative
$\partial_{\hat{i}}$ by parts, and obtaining a boundary term, are
\begin{subequations}
  \begin{align}
   - \int d^3 Y\,n^{\hat{i}} \tilde{T}^{(0)\hat{k}\hat{j}} \delta_{\hat{i}\hat{j}} = 0 \hspace{2cm}  &(m=1,N=0), \label{eq:sespacecompleading10}\\
- \int d^3 Y\,\tilde{T}^{(0)\hat{k}\hat{i}} = 0 \hspace{2cm} &(m=1,N=1),\\
- \int d^3 Y\, R\, \tilde{T}^{(0)\hat{k}\hat{l}} -  \int d^3 Y\, n^{\hat{l}} n^{\hat{i}} R\, \tilde{T}^{(0)\hat{k}\hat{j}} \delta_{\hat{i}\hat{j}} -  \tfrac{1}{6} \left(\tilde{\mathcal{J}}^{(0)\hat{0}}\right)^2 \delta^{\hat{k}\hat{l}} = 0\hspace{2cm}  &(m=2,N=1), \label{eq:leadingOrderNonZeroBoundary}\\
- \int d^3 Y\,n^{\hat{j}} R\, \tilde{T}^{(0)\hat{k}\hat{i}} -  \int d^3 Y\,n^{\hat{i}} R\, \tilde{T}^{(0)\hat{k}\hat{j}} =0\hspace{2cm} &(m=2,N=2).
\end{align}
\end{subequations}
The boundary terms can be evaluated using Eqs. (\ref{eq:retardedField}),(\ref{eq:scaledSelfField}),(\ref{eq:fieldSE}), and (\ref{eq:rescaledSE}) and are nonzero only in (\ref{eq:leadingOrderNonZeroBoundary}).

The first moments of the time component (\ref{eq:timeseIntegration}) yield
\begin{subequations}
  \begin{align}
- \int d^3 Y\,n^{\hat{i}} \tilde{T}^{(0)\hat{j}\hat{0}} \delta_{\hat{i}\hat{j}}   =0 \hspace{2cm} &(m=1,N=0) ,\\
 - \int d^3 Y\,\tilde{T}^{(0)\hat{i}\hat{0}}=0 \hspace{2cm} &(m=1,N=1) , \label{eq:setimecompleading11}\\
- \int d^3 Y \,R\, \tilde{T}^{(0)\hat{k}\hat{0}} -  \int d^3 Y \,n^{\hat{k}}n^{\hat{i}} R\, \tilde{T}^{(0)\hat{j}\hat{0}} \delta_{\hat{i}\hat{j}} =0 \hspace{2cm} &(m=2,N=1)  , \label{eq:setimecompleading21}\\
- \int d^3 Y\,n^{\hat{j}} R\, \tilde{T}^{(0)\hat{i}\hat{0}} -  \int d^3 Y\,n^{\hat{i}} R\, \tilde{T}^{(0)\hat{j}\hat{0}}=0 \hspace{2cm} &(m=2,N=2)  ,\\
\end{align}
\end{subequations}
It follows from (\ref{eq:sespacecompleading10}), (\ref{eq:setimecompleading11}), and (\ref{eq:retardedmomentum}) that
\begin{equation} \label{eq:leadingmomentum}
  \tilde{P}^{\mu} = \tilde{m} u^\mu + \mathcal{O}(\lambda).
\end{equation}

The first  moments of (\ref{eq:framecurrentcons}) yield
\begin{subequations}
  \begin{align}
    -\int d^3Y n^{\hat{i}} \tilde{j}^{\hat{j}} \delta_{\hat{i} \hat{j}} = 0\hspace{2cm}&(m=1,N=0), \label{eq:leadingOrderChargecurrent10}\\
  - \int d^3 Y\, j^{(0)\hat{i}} =0 \hspace{2cm}&(m=1,N=1),\label{eq:leadingOrderChargecurrent11}\\
  - \int d^3 Y j^{(0)\hat{j}} n^{\hat{i}} R\, -  \int d^3 Y j^{(0)\hat{i}} n^{\hat{j}} R\,   =0 \hspace{2cm}&(m=2,N=2),\\
  - \int d^3 Y j^{(0)\hat{k}} R\,^2 - 2 \int d^3 Y j^{(0)\hat{i}} n^{\hat{k}}n^{\hat{j}} R\,^2 \delta_{\hat{i}\hat{j}}   =0 \hspace{2cm}&(m=3,N=1),\\
- \int d^3 Y j^{(0)\hat{k}} n^{\hat{i}}n^{\hat{j}} R^2 -  \int d^3 Y j^{(0)\hat{j}} n^{\hat{i}}n^{\hat{k}} R^2 -  \int d^3 Y j^{(0)\hat{i}} n^{\hat{j}}n^{\hat{k}} R^2  =0 \hspace{2cm}&(m=3,N=3).
\end{align}
\end{subequations}
It follows from
Eqs. (\ref{eq:leadingOrderChargecurrent10}),(\ref{eq:leadingOrderChargecurrent11}),
and (\ref{eq:retardedCharge}) that
\begin{equation} \label{eq:leadingcharge}
  \tilde{\mathcal{J}}^{\hat{a}} = \tilde{q} u^{\hat{a}} + \mathcal{O}(\lambda).
\end{equation}

This process may be continued to each higher order in $\lambda$.  At first order
in $\lambda$, from the $(m=0,N=0)$ piece of (\ref{eq:spatialseIntegration}) we obtain
\begin{align} \label{eq:firstorderforceraw}
0=&  F^{(\text{ext})\hat{k}\hat{0}} \int d^3 Y \,\tilde{j}^{(0)\hat{0}} + a^{(0) \hat{k}} \int d^3 Y \,\tilde{T}^{(0)\hat{0}\hat{0}} -  F^{(\text{ext})\hat{k}\hat{j}} \int d^3 Y \, \tilde{j}^{(0)\hat{i}} \delta_{\hat{i}\hat{j}} \notag\\&-  a^{(0)\hat{k}} \int d^3 Y \, n^{\hat{i}} \tilde{T}^{(0)\hat{j}\hat{0}} \delta_{\hat{i}\hat{j}}  + \int d^3 Y\,\tilde{T}^{(0)\hat{k}\hat{0}}{}_{,0} -  \delta_{\hat{i}\hat{j}} \int d^3 Y \,n^{\hat{i}} \tilde{T}^{(0)\hat{k}\hat{j}}{}_{,0},
\end{align}
where the external field is evaluated on the worldline. Combining
(\ref{eq:firstorderforceraw}) with
(\ref{eq:retardedmomentumspin}),(\ref{eq:retardedmoments}),(\ref{eq:sespacecompleading10}),(\ref{eq:setimecompleading21}),
and (\ref{eq:leadingOrderChargecurrent11}) gives,
\begin{equation} \label{eq:leadingprelorentzforce}
  \partial_{\tau} \tilde{P}^{(0)\hat{i}} + \tilde{P}^{(0)\hat{0}} a^{(0)\hat{i}} = D_\tau \tilde{P}^{(0)\hat{i}}  = - F^{(\text{ext})\hat{i} \hat{0}} \tilde{\mathcal{J}}^{(0)\hat{0}}.
\end{equation}
Similarly, the $\mathcal{O}(\lambda)$ piece of the $(m=0,N=0)$ piece of
Eq.(\ref{eq:timeseIntegration}) together with (\ref{eq:leadingcharge}) and
(\ref{eq:leadingmomentum}) gives
\begin{equation}
  \partial_\tau \tilde{m} = \mathcal{O}(\lambda).
\end{equation}
Combining this with (\ref{eq:leadingprelorentzforce}) gives
\begin{equation}
  \tilde{m} a^{ (0)\hat{i}} = -F^{(\text{ext})\hat{i} \hat{0}} \tilde{\mathcal{J}}^{(0)\hat{0}},
\end{equation}
the Lorentz force law.

This procedure may be extended to higher moments, and to higher orders in
perturbation theory to yield the self force expressions in
Secs. \ref{sec:firstOrderDerivation}-\ref{sec:2ndorderderive}, giving the final results
presented in Sec. \ref{sec:finalresultsRestricted}.

The computation of the set of equations (\ref{eq:integrationSet}) was automated,
using the Mathematica computer algebra software. The notebook used to compute
the self force can be found at \cite{moxonWebspace}. The equations we present
take advantage of the worldline-based tetrads in the retarded coordinates to
re-assemble a covariant form for the laws of motion, so retarded coordinates
appear nowhere in our final results in section \ref{sec:laws}. The hierarchy of
equations (\ref{eq:integrationSet}) is similar to that used by GHW, except that
they use integrals over spacelike hypersurfaces

\subsection{First order laws of motion: Abraham-Lorentz-Dirac} \label{sec:firstOrderDerivation}

\subsubsection{Derivation of law of motion}

To derive the first order laws of motion, we expand the scaled field equations
(\ref{eq:frameSEcons}) and (\ref{eq:framecurrentcons}) to second order in
$\lambda$. We will need to use the spin supplementary condition for the first
order laws of motion, so we'll present first the leading self-torque, and we
will derive the required spin renormalization (\ref{eq:restrictedSpin}) from the
leading order self-torque.

We first compute the component of the bare momentum orthogonal to the worldline
through $\mathcal{O}(\lambda)$ by combining the $(m=1,N=0)$ piece of
(\ref{eq:spatialseIntegration}) at $\mathcal{O}(\lambda)$ with the $(m=1,N=1)$
piece of (\ref{eq:timeseIntegration}), together with
(\ref{eq:retardedmomentumspin}), (\ref{eq:retardedmoments}). The result is
\begin{align} \label{eq:bareMom1}
  \tilde{P}^\eta \mathcal{P}_\eta{}^\mu =& - \lambda\frac{2}{3} \tilde{q}^2 a^\mu
  +\lambda \mathcal{P}^\mu{}_\eta D_\tau \tilde{S}^{\eta \nu} u_\nu + 2 \lambda
  \mathcal{P}^\mu{}_\eta F^{(\text{ext})[\eta}{}_\lambda \tilde{Q}^{\lambda | \nu]} u_\nu +
  \mathcal{O}(\lambda^2).
\end{align}
Here we have converted from equations involving tetrad components to covariant
equations, by using the fact that derivatives with respect to $\tau$ of tetrad
components evaluate on the worldline can be converted to covariant Fermi
derivatives $D_F/d\tau$ \cite{GHW}, defined for any vector $v^\mu$ by
\begin{equation}
  \frac{D_F}{d\tau} v^{\mu} = \frac{D}{d\tau} v^{\mu} + (a^\mu u^\nu - a^\nu
  u^\mu)v_{\nu}.
\end{equation}
We also note that Eq. (\ref{eq:bareMom1}) could equivalently have been derived
directly from (\ref{eq:exactMomentum}) instead of by taking moments of the field
equation.

We next compute the first covariant derivative of both the bare momentum and the
bare spin through $\mathcal{O}(\lambda^2)$. The covariant derivative of the bare
momentum is obtained from the $(m=0,N=0)$ moment of the equations
(\ref{eq:spatialseIntegration},\ref{eq:timeseIntegration}) and the covariant
derivative of the spin is obtained from the antisymmetrized moment
(\ref{eq:spatialseIntegration}) $(m=1,N=1)$.
\begin{subequations} 
  \begin{align}
D_\tau \tilde{P}^\lambda =&  F^{(\text{ext}) \lambda \mu} \tilde{\mathcal{J}}_\mu +\lambda F^{(\text{ext}) \lambda}{}_{\mu;\nu} \tilde{Q}^{\mu \nu}- \lambda \frac{2}{3} \tilde{q}^2 a_\nu a^\nu u^\lambda  + \mathcal{O}(\lambda^2),
   \label{eq:bareCovDMom1}\\
 D_\tau \tilde{S}^{\mu \nu} =& F^{(\text{ext}) [\mu}{}_\lambda
    \tilde{Q}^{\lambda \nu]} + \mathcal{O}(\lambda). \label{eq:bareTorque1}
  \end{align}
\end{subequations}

We also expand the rest mass, which contains no new correction at this order, by
combining (\ref{eq:leadingmomentum}), (\ref{eq:scaledmdef}), and
(\ref{eq:bareMom1}). The result is
\begin{equation} \label{eq:bareMass1}
  \tilde{m} = -\tilde{P}^\mu u_\mu + \mathcal{O}(\lambda^2).
\end{equation}

At this point, we have imposed no spin supplementary condition, so these
equations are entirely general\footnote{To this order in perturbation theory,
  and provided the definitions given in section \ref{sec:pointparticle}.}, but do
not describe the evolution of a worldline. To compute the center of
mass acceleration, we use the spin supplementary condition (\ref{eq:spinsup}),
which reduces at this order to, from Eq. (\ref{eq:restrictedSpin})
\begin{equation}
\tilde{S}^{\mu \nu} u_\nu  =\mathcal{O}(\lambda).
\end{equation}
Combining Eqs. (\ref{eq:bareMom1})-(\ref{eq:bareMass1}), we deduce the acceleration
and evolution of the rest mass:
\begin{subequations}
  \begin{align}
   a^{\sigma}\tilde{m} =& \mathcal{P}^\sigma{}_\mu \bigg[ F^{(\text{ext})\mu
       \nu} \tilde{\mathcal{J}}_\nu +\lambda F^{(\text{ext}) \mu}{}_{\lambda;\nu}
     \tilde{Q}^{\lambda \nu} + \lambda \tfrac{2}{3} \tilde{q}^2 D_\tau a^\mu + D_\tau
     \left( a_{\lambda} \tilde{S}^{ \mu \lambda} + u_\nu F^{(\text{ext})
       [\nu|}{}_{\rho} \tilde{Q}^{\rho |\mu]}\right) \bigg] +
   \mathcal{O}(\lambda^2), \label{eq:o1enm} \\ D_\tau \tilde{m} =& -u_\mu F^{(\text{ext})\mu
     \lambda} \tilde{\mathcal{J}}_\lambda + u_\mu F^{(\text{ext})
     \mu}{}_{\lambda;\eta} \tilde{Q}^{\lambda \eta} - 2 a_\eta
   F^{(\text{ext})[\eta}{}_\lambda \tilde{Q}^{\lambda | \nu]} u_\nu +
   \mathcal{O}(\lambda^2).\label{eq:massD1}
  \end{align}
\end{subequations}
In addition, we find from the $(m=0,N=0)$ component of the charge conservation
equation (\ref{eq:currentIntegration}) at $\mathcal{O}(\lambda)$ that
\begin{equation}
  D_\tau \tilde{q} = \mathcal{O}(\lambda),
\end{equation}
consistent with the fact that charge is conserved to all orders. From the
$(m=1,N=0)$ and $(m=1,N=1)$ pieces together with Eq. (\ref{eq:retardedmoments}), we
find the expression for the charge moment to this order,
\begin{align} \label{eq:chargemomentval1}
  \tilde{\mathcal{J}}^{\mu} =& \tilde{q} u^\mu + \lambda u_\nu D_\tau \tilde{Q}^{\mu \nu} - \lambda
  a^\mu u_\nu u_\lambda \tilde{Q}^{\nu \lambda} - \lambda \mathcal{P}^\mu{}_\nu u_\lambda D_\tau
  \tilde{Q}^{\lambda \nu} + \mathcal{O}(\lambda^2).
\end{align}

We next rewrite our results (\ref{eq:bareCovDMom1}),(\ref{eq:o1enm}), and
(\ref{eq:massD1}) in terms of the projected, renormalized body parameters~(\ref{eq:restrictedMass})~-~(\ref{eq:restrictedDipole}) and eliminate
$\tilde{\mathcal{J}}^\mu$ using (\ref{eq:chargemomentval1}). This yields the results
(\ref{eq:GHWreproduce}) and the leading piece of (\ref{eq:restrictedTorque})
given in the previous section.

\subsubsection{Consistency check using the Harte equation of motion} \label{sec:GHWtotalderiv}

We now preform the consistency check described in Sec. \ref{sec:Hartecheck}. The
radiative self field $F_R^{\mu \nu}$ in Eq. (\ref{eq:diffconfirmation}) is given
by \cite{HarteExact} and \cite{Poisson}, for which the only non-vanishing
component is
\begin{equation}
   \hat{F}^{\eta \sigma} \mathcal{P}_{\eta}{}^\lambda u_\sigma = F^{(\text{ext})
     \eta \sigma} \mathcal{P}_{\eta}{}^\lambda u_\sigma + \lambda \frac{2}{3}
   \tilde{q} D_\tau a^{ \eta} \mathcal{P}_{\eta}{}^\lambda+
   \mathcal{O}(\lambda^2).
    \label{eq:o1momenm-fieldrenorm}
\end{equation}
The self stress energy tensor can also be computed from
Eq. (\ref{eq:retardedField}); see also Eq. (120) of GHW. Substituting into
Eq. (\ref{eq:diffconfirmation}) gives that,
\begin{equation}
  D_\tau P^\mu_H - D_\tau P_B^\mu = \lambda^2 D_\tau \left(\frac{2}{3}\tilde{q}^2 a^\mu\right) + \mathcal{O}(\lambda^3),
\end{equation}
and so the right hand side is indeed a total derivative, as required.

\subsection{New result: second order laws of motion} \label{sec:2ndorderderive}
\label{sec:finalresultsbare}

\subsubsection{Derivation of laws of motion}

The derivation at second order parallels the derivation given above at first
order. We follow the same steps as before, to one higher order in $\lambda$.
First, we derive the bare momentum orthogonal to the worldline from moments
$(m=1,N=0)$ of (\ref{eq:spatialseIntegration}) and $(m=1,N=1)$ of
(\ref{eq:timeseIntegration}) through second order. After simplifying according
to equations obtained from the full set of moments from $\mathcal{O}(\lambda^2)$
equations, we obtain
\begin{align} \label{eq:secondOrderBareMomentum}
  \tilde{P}^{\kappa} \mathcal{P}_\kappa{}^\mu = \lambda \mathcal{P}^\mu{}_\kappa
  \bigg[& - \frac{2}{3} \tilde{q}^2 a^\kappa + D_\tau \tilde{S}^{\kappa \nu}
    u_\nu + 2 F^{(\text{ext})[\kappa}{}_\lambda \tilde{Q}^{\lambda | \nu]} u_\nu + 4 \lambda
    \tilde{Q}^{\lambda \nu [\kappa} F^{(\text{ext})\sigma]}{}_{\lambda;\nu}
    u_\sigma + 3\lambda \tilde{q} a_{\nu} D_{\tau}\tilde{Q}^{\nu\kappa}
    \nonumber \\ & - \tfrac{1}{3}\lambda \tilde{q} a_{\nu}
    D_{\tau}\tilde{Q}^{\kappa\nu}+ \lambda \tilde{q} \tilde{Q}^{\kappa \nu}
    \left(\tfrac{1}{3} u_\nu a_\sigma a^\sigma - \tfrac{2}{3} D_\tau
    a_\nu\right) + \tfrac{4}{3} \lambda \tilde{q} D_{\tau}a^{\kappa}
    \tilde{Q}^{\nu \lambda} u_\nu u_\lambda + \tfrac{1}{5}\lambda \tilde{q}
    a_{\nu} a^{\nu} \tilde{Q}^{\lambda \kappa} u_\lambda \notag\\ &+
    \tfrac{4}{3}\lambda \tilde{q} D_{\tau}^2 \tilde{Q}^{\nu \kappa} u_\nu +
    \lambda \tilde{q} a^{\kappa}\left(\tfrac{8}{3} D_{\tau}\tilde{Q}^{\nu
      \lambda} u_\nu u_\lambda + \tfrac{7}{5} a_{\nu} \tilde{Q}^{\lambda \nu}
    u_\lambda + 3 a_{\nu} \tilde{Q}^{\nu\lambda}u_\lambda \right) \bigg] +
  \mathcal{O}(\lambda^3).
\end{align}

The higher-order moments fix also the first covariant derivatives of the bare moments. The first derivative of the bare momentum arises from the $(m=0,N=0)$ moment of the equations (\ref{eq:spatialseIntegration},\ref{eq:timeseIntegration}), and subsequent simplifications from $\mathcal{O}(\lambda^2)$ moments, and takes the value
\begin{subequations} \label{eq:CovDMom2}
\begin{align} \label{eq:secondOrderDbareMoment}
    \mathcal{P}^\sigma{}_\kappa D_\tau \tilde{P}^{\kappa} =\mathcal{P}^\sigma{}_\kappa \bigg[& F^{\kappa \nu} \tilde{\mathcal{J}}_\mu  + \lambda F^{(\text{ext}) \kappa}{}_{\lambda;\nu} \tilde{Q}^{\lambda \nu}  + \tfrac{1}{2}\lambda^2  F^{(\text{ext}) \kappa}{}_{\nu;\lambda \sigma} \tilde{Q}^{\nu \lambda \sigma}  \notag\\
      &+ \lambda^2 \tilde{q}  D_\tau a^{\kappa} a_{\mu}  u_\nu \left(  \tfrac{1}{5}   \tilde{Q}^{\nu \mu} - \tfrac{1}{3}  \tilde{Q}^{\mu \nu}\right)   + \lambda^2 q a_{\mu} D_\tau a^{\mu} u_\nu \left( \tfrac{1}{5}  \tilde{Q}^{\nu\kappa} +  \tfrac{1}{3} \lambda^2 \tilde{Q}^{\kappa\nu} \right) \notag\\& + \lambda^2 \tilde{q} a_{\mu} a^{\mu} \left(\tfrac{4}{15} D_{\tau}\tilde{Q}^{\nu\kappa}u_\nu   + \tfrac{2}{3} \lambda^2 D_{\tau}\tilde{Q}^{\kappa\nu}u_\nu   + \tfrac{4}{15}  a_{\nu}  \tilde{Q}^{\nu \kappa} \right)   +  \tfrac{2}{3} \lambda^2 \tilde{q} a_{\mu}  D_{\tau}^2 \tilde{Q}^{[\kappa\mu]} 
    \nonumber \\
    & + \lambda^2 \tilde{q} a^{\kappa}  \left(\tfrac{8}{15} a_{\mu} D_{\tau}\tilde{Q}^{\nu\mu} u_\nu - \tfrac{2}{3}  a_{\mu}  D_{\tau}\tilde{Q}^{\mu\nu}u_\nu  + \tfrac{4}{5} a_{\mu} a^{\mu} \tilde{Q}^{\nu \lambda} u_\nu u_\lambda  - \tfrac{2}{15} D_{\tau}a_{\mu} \mathcal{P}^\mu{}_\nu \tilde{Q}^{\lambda \nu} u_\lambda  \right)   \bigg] \notag\\&+ \mathcal{O}(\lambda^3)\\
    u_\mu D_\tau \tilde{P}^\mu  =&  u_\nu F^{(\text{ext}) \nu \mu} \tilde{\mathcal{J}}_\mu  + \tfrac{2}{3}\lambda \tilde{q}^2 a_{\mu} a^{\mu}   +\lambda  u_\mu  F^{(\text{ext})\mu}{}_{\nu;\lambda} \tilde{Q}^{\nu \lambda}  +  \tfrac{1}{2} \lambda^2 u_\mu F^{(\text{ext}) \mu}{}_{\nu; \lambda; \sigma} \tilde{Q}^{\nu \lambda \sigma}  \nonumber \\
  & - \lambda^2 \tilde{q} a_{\mu} a^{\mu} \left( \tfrac{8}{3}  D_{\tau}\tilde{Q}^{\nu \lambda} u_\nu u_\lambda +  \tfrac{22}{15}  a_{\nu} \tilde{Q}^{\lambda\nu}u_\lambda  +  \tfrac{8}{3} a_{\nu} \tilde{Q}^{\nu \lambda} u_\lambda \right)   - \tfrac{4}{3}  \lambda^2 \tilde{q} a_{\mu} D_{\tau}a^{\mu} \tilde{Q}^{\nu \lambda}u_\nu u_\lambda   \notag\\&-  \tfrac{4}{3} \lambda^2 \tilde{q} a_{\mu}  D_{\tau}^2 \tilde{Q}^{\nu\mu} u_\nu - \lambda^2 \tfrac{2}{3} \tilde{q} a_{\mu} D_{\tau}a_\nu  \tilde{Q}^{\nu \mu} + \mathcal{O}(\lambda^4).
\end{align}
\end{subequations}
The torque is computed from the antisymmetric part $(m=1,N=1)$ of (\ref{eq:spatialseIntegration}) and simplifications from $\mathcal{O}(\lambda^2)$ equations,
\begin{align}
  D_\tau \tilde{S}^{\nu \lambda} \mathcal{P}_\nu{}^\kappa \mathcal{P}_{\lambda}{}^\mu =& 2 F^{(\text{ext}) [\kappa|}{}_\nu \tilde{Q}^{\nu |\mu]} + 2 \lambda F^{(\text{ext}) [\kappa|}{}_{\nu ; \lambda} \tilde{Q}^{\nu \lambda |\mu]}
 + 2 \lambda \tilde{q} a^{[\kappa} \mathcal{P}^{\mu]}{}_\lambda \left(\tfrac{1}{3} D_\tau \tilde{Q}^{\nu \lambda}u_\nu + \tfrac{2}{3} D_\tau \tilde{Q}^{\lambda \nu} u_\nu \right) \notag\\& + \tfrac{2}{3} \lambda \tilde{q} D_\tau a^{[\kappa} \tilde{Q}^{\mu] \nu} u_\nu - \tfrac{2}{3} \lambda \tilde{q} D_\tau^2 \tilde{Q}^{[\kappa \mu]}  + \mathcal{O}(\lambda^2).
\end{align}

The rest mass is derived by expanding \ref{eq:scaledmdef}, using  the bare momentum (\ref{eq:bareMom1}). This gives
\begin{align} \label{eq:selfForcePart1}
 \tilde{m} + \tilde{P}^\mu u_\mu = \lambda^2 \frac{1}{\tilde{m}}\bigg(& -\frac{2}{9} \tilde{q}^4 a_\mu a^\mu + \frac{8}{3} \tilde{q}^2 a_\mu F^{(\text{ext}) [\mu|}{}_\nu \tilde{Q}^{\nu |\lambda]} u_\lambda-  \frac{1}{2} a_{\mu} a_{\nu} \tilde{S}^{\mu}{}_\kappa \tilde{S}^{\nu \kappa}  \nonumber \\
 & -   a_{\nu} F^{(\text{ext})}{}_{\mu}{}^{[\lambda|} \tilde{Q}^{\mu|\eta]} \tilde{S}^{\nu}{}_{\lambda} u_\eta  -  \frac{1}{2} F^{(\text{ext})}{}_{\kappa}{}^{[\lambda|} F^{(\text{ext})\mu}{}_{[\nu|} \tilde{Q}^{\kappa|\sigma]} \tilde{Q}_{\mu|\eta]} u_\sigma u^\eta  \mathcal{P}^\nu{}_\lambda  \bigg)  + \mathcal{O}(\lambda^3).
\end{align}

Similarly, we derive the charge moment through second order using the $(m=1,N=0)$
and $(m=1,N=1)$ pieces of Eq.(\ref{eq:currentIntegration}) at
$\mathcal{O}(\lambda^2)$. The result is
\begin{align} \label{eq:secondOrderChargeMoment}
  \tilde{\mathcal{J}}^{\mu} =& \tilde{q} u^\mu + \lambda u_\nu D_\tau \tilde{Q}^{\mu \nu} - \lambda a^\mu u_\nu u_\lambda Q^{\nu \lambda} - \lambda \mathcal{P}^\mu{}_\nu u_\lambda D_\tau \tilde{Q}^{\lambda \nu} - \tfrac{1}{2} \lambda^2 D_{\tau}a^{\mu } \tilde{Q}^{\nu \lambda \rho }u_\nu u_\lambda u_\rho -  \tfrac{1}{2} \lambda^2 D_{\tau}^2 \tilde{Q}^{\mu \nu \lambda} u_\nu u_\lambda  \notag\\
  &  - \lambda^2 a^\mu \left(  \tfrac{3}{2} a_{\nu}  \tilde{Q}^{(\nu\lambda \rho)}u_\lambda u_\rho + \tfrac{3}{2} D_{\tau} \tilde{Q}^{\nu \lambda \rho} u_\nu u_\lambda u_\rho \right) - \lambda^2 \mathcal{P}^\mu{}_\nu \left(3  D_{\tau}\tilde{Q}^{(\nu \lambda \rho)} a_{\lambda} u_\rho +   D_{\tau}^2 \tilde{Q}^{\lambda \nu \rho} u_\lambda u_\rho  \right) +  \mathcal{O}(\lambda^3).
\end{align}

Finally, to evaluate the explicit equations of motion for the worldline and for
the evolution of the rest mass, we use the following rescaled versions of the
general identities (\ref{eq:lawofmotionidentity}):
\begin{subequations}  \label{eq:genSelfForce-1}
  \begin{align}
   \tilde{m} a^\kappa =& a^\kappa \left(\tilde{m} + \tilde{P}^\mu u_\mu \right) + \mathcal{P}^\kappa{}_\lambda D_\tau \tilde{P}^\lambda - \mathcal{P}^\kappa{}_\nu D_\tau \left(\mathcal{P}^\nu{}_\lambda \tilde{P}^\lambda\right),  \\
    D_\tau \tilde{m} =& D_\tau \left(\tilde{m} + \tilde{P}^\mu u_\mu\right) - u_\mu D_\tau \tilde{P}^\mu - a_\mu \tilde{P}^\mu,
  \end{align}
\end{subequations}
One can think of the first and third terms in each of (\ref{eq:genSelfForce-1})
as representing the effect of hidden momentum, that is, the component of
momentum perpendicular to $\vec{u}$. By substituting the results
(\ref{eq:secondOrderBareMomentum})-(\ref{eq:selfForcePart1}) and
(\ref{eq:secondOrderChargeMoment}) into the general identity
(\ref{eq:genSelfForce-1}), making use of the spin supplementary condition
(\ref{eq:spinsup}), and eliminating the body parameters in terms of the
renormalized projected body parameters
(\ref{eq:restrictedMass})-(\ref{eq:restrictedQuadrupole}), we finally arrive at
the second order equations of motion
(\ref{eq:secondorderstart})-(\ref{eq:diquadMdotSecond}).

\subsubsection{Consistency check using the Harte equation of motion} \label{sec:secondordertotalderiv}

We turn now to the consistency check described in Sec. \ref{sec:Hartecheck}. We
first compute the regular self field through second order. The expansions
(\ref{eq:retardedField}) we use to derive these expressions are expanded
asymptotically at large $R$. However, taking the difference between the retarded
and advanced fields in the multipole expansion, re-expanded at small $R$ will
yield the regular field. This procedure can be thought of as obtaining an
asymptotic form for the fields, then replacing the extended source with a
pointlike source for the purposes of computing the regular field. Since the
regular field should depend only on the standard multipoles of the body, the
regular field should be indistinguishable for the extended and replaced pointlike
body. This procedure and argument are analogous to that used by Pound
\cite{poundgauge} for the gravitational case.

The result in terms of the tetrad components and retarded coordinates from
\ref{sec:computationSteps} is
\begin{align}
  F_{R\, \hat{k} \hat{0}} =&  \tfrac{2}{3} \lambda D_{\tau}a^{\hat{i}} \tilde{q} \delta_{\hat{k}\hat{i}} - \tfrac{2}{3} \lambda^2 a_{\hat{i}} a^{\hat{i}} a_{\hat{k}} \tilde{q} R -  \tfrac{2}{3} \lambda^2 a_{\hat{k}} D_{\tau}a^{\hat{i}} \tilde{q} n^{\hat{j}} R \delta_{\hat{i}\hat{j}}  + \tfrac{2}{3} \lambda^2 D_{\tau}{}^2 a^{\hat{i}} \tilde{q} R \delta_{\hat{k}\hat{i}} \nonumber \\
  & -  \tfrac{4}{3} \lambda^2 a_{\hat{i}} D_{\tau}a^{\hat{j}} \tilde{q} n^{\hat{i}} R \delta_{\hat{k}\hat{j}} + \tfrac{2}{3} \lambda^2 a_{\hat{i}} D_{\tau}a^{\hat{i}} \tilde{q} n^{\hat{j}} R \delta_{\hat{k}\hat{j}} -  \tfrac{2}{3} \lambda^2 a_{\hat{i}} D_{\tau}a^{\hat{i}} \tilde{Q}^{\hat{0}\hat{j}} \delta_{\hat{k}\hat{j}} -  \tfrac{2}{3} \lambda^2 D_{\tau}{}^2 a^{\hat{i}} \tilde{Q}^{\hat{j}\hat{l}} \delta_{\hat{i}\hat{l}} \delta_{\hat{k}\hat{j}} \nonumber \\
  & + \tfrac{1}{3} \lambda^2 a_{\hat{i}} a^{\hat{i}} a_{\hat{j}} \tilde{Q}^{\hat{j}\hat{l}} \delta_{\hat{k}\hat{l}} + \tfrac{2}{3} \lambda^2 a_{\hat{i}} a^{\hat{i}} a_{\hat{j}} \tilde{Q}^{\hat{l}\hat{j}} \delta_{\hat{k}\hat{l}} -  \tfrac{1}{3} \lambda^2 D_{\tau}{}^2 a^{\hat{i}} \tilde{Q}^{\hat{j}\hat{l}} \delta_{\hat{i}\hat{j}} \delta_{\hat{k}\hat{l}} -  \tfrac{2}{3} \lambda^2 a_{\hat{i}} a^{\hat{i}} \delta_{\hat{k}\hat{j}} \partial_\tau \tilde{Q}^{\hat{0}\hat{j}} \nonumber \\
  & -  \tfrac{5}{3} \lambda^2 D_{\tau}a^{\hat{i}} \delta_{\hat{i}\hat{l}} \delta_{\hat{k}\hat{j}} \partial_\tau \tilde{Q}^{\hat{j}\hat{l}} -  \lambda^2 D_{\tau}a^{\hat{i}} \delta_{\hat{i}\hat{l}} \delta_{\hat{k}\hat{j}} \partial_\tau \tilde{Q}^{\hat{l}\hat{j}} -  \lambda^2 a_{\hat{i}} \delta_{\hat{k}\hat{j}} \partial_\tau{}^2 \tilde{Q}^{\hat{i}\hat{j}} -  \lambda^2 a_{\hat{i}} \delta_{\hat{k}\hat{j}}  \partial_\tau{}^2 \tilde{Q}^{\hat{j}\hat{i}} \nonumber \\
  & + \tfrac{2}{3} \lambda^2 \delta_{\hat{k}\hat{i}} \partial_\tau{}^3 \tilde{Q}^{\hat{0}\hat{i}} + \mathcal{O}(\lambda^3),
\end{align}
and
\begin{align}
  F_{R\, \hat{k} \hat{j}} =& \tfrac{2}{3} \lambda^2 a_{\hat{k}} D_{\tau}a^{\hat{i}} \tilde{q} R \delta_{\hat{j}\hat{i}} -  \tfrac{1}{3} \lambda^2 a_{\hat{i}} a^{\hat{i}} a_{\hat{k}} \tilde{q} n^{\hat{l}} R \delta_{\hat{j}\hat{l}} + \tfrac{1}{3} \lambda^2 a_{\hat{i}} a^{\hat{i}} a_{\hat{k}} \tilde{Q}^{\hat{0}\hat{l}} \delta_{\hat{j}\hat{l}} -  \tfrac{2}{3} \lambda^2 a_{\hat{k}} D_{\tau}a^{\hat{i}} \tilde{Q}^{\hat{l}\hat{m}} \delta_{\hat{i}\hat{m}} \delta_{\hat{j}\hat{l}} \nonumber \\
  & -  \tfrac{2}{3} \lambda^2 a_{\hat{k}} D_{\tau}a^{\hat{i}} \tilde{Q}^{\hat{l}\hat{m}} \delta_{\hat{i}\hat{l}} \delta_{\hat{j}\hat{m}} -  \tfrac{2}{3} \lambda^2 a_{\hat{j}} D_{\tau}a^{\hat{i}} \tilde{q} R \delta_{\hat{k}\hat{i}} + \tfrac{1}{3} \lambda^2 D_{\tau}{}^2 a^{\hat{i}} \tilde{q} n^{\hat{l}} R \delta_{\hat{j}\hat{l}} \delta_{\hat{k}\hat{i}} -  \tfrac{1}{3} \lambda^2 D_{\tau}{}^2 a^{\hat{i}} \tilde{Q}^{\hat{0}\hat{l}} \delta_{\hat{j}\hat{l}} \delta_{\hat{k}\hat{i}} \nonumber \\
  & + \tfrac{1}{3} \lambda^2 a_{\hat{i}} a^{\hat{i}} a_{\hat{j}} \tilde{q} n^{\hat{l}} R \delta_{\hat{k}\hat{l}} -  \tfrac{1}{3} \lambda^2 a_{\hat{i}} a^{\hat{i}} a_{\hat{j}} \tilde{Q}^{\hat{0}\hat{l}} \delta_{\hat{k}\hat{l}} + \tfrac{2}{3} \lambda^2 a_{\hat{j}} D_{\tau}a^{\hat{i}} \tilde{Q}^{\hat{l}\hat{m}} \delta_{\hat{i}\hat{m}} \delta_{\hat{k}\hat{l}} -  \tfrac{1}{3} \lambda^2 D_{\tau}{}^2  a^{\hat{i}} \tilde{q} n^{\hat{l}} R \delta_{\hat{j}\hat{i}} \delta_{\hat{k}\hat{l}} \nonumber \\
  & + \tfrac{1}{3} \lambda^2 D_{\tau}{}^2 a^{\hat{i}} \tilde{Q}^{\hat{0}\hat{l}} \delta_{\hat{j}\hat{i}} \delta_{\hat{k}\hat{l}} -  \tfrac{1}{3} \lambda^2 a_{\hat{i}} D_{\tau}a^{\hat{l}} \tilde{Q}^{\hat{i}\hat{m}} \delta_{\hat{j}\hat{m}} \delta_{\hat{k}\hat{l}} + \tfrac{1}{3} \lambda^2 a_{\hat{i}} D_{\tau}a^{\hat{i}} \tilde{Q}^{\hat{l}\hat{m}} \delta_{\hat{j}\hat{m}} \delta_{\hat{k}\hat{l}} \nonumber \\
  & -  \tfrac{1}{3} \lambda^2 a_{\hat{i}} D_{\tau}a^{\hat{l}} \tilde{Q}^{\hat{m}\hat{i}} \delta_{\hat{j}\hat{m}} \delta_{\hat{k}\hat{l}} + \tfrac{2}{3} \lambda^2 a_{\hat{j}} D_{\tau}a^{\hat{i}} \tilde{Q}^{\hat{l}\hat{m}} \delta_{\hat{i}\hat{l}} \delta_{\hat{k}\hat{m}} + \tfrac{1}{3} \lambda^2 a_{\hat{i}} D_{\tau}a^{\hat{l}} \tilde{Q}^{\hat{i}\hat{m}} \delta_{\hat{j}\hat{l}} \delta_{\hat{k}\hat{m}} \nonumber \\
  & -  \tfrac{1}{3} \lambda^2 a_{\hat{i}} D_{\tau}a^{\hat{i}} \tilde{Q}^{\hat{l}\hat{m}} \delta_{\hat{j}\hat{l}} \delta_{\hat{k}\hat{m}} + \tfrac{1}{3} \lambda^2 a_{\hat{i}} D_{\tau}a^{\hat{l}} \tilde{Q}^{\hat{m}\hat{i}} \delta_{\hat{j}\hat{l}} \delta_{\hat{k}\hat{m}} -  \tfrac{2}{3} \lambda^2 D_{\tau}a^{\hat{i}} \delta_{\hat{j}\hat{l}} \delta_{\hat{k}\hat{i}} \partial_\tau \tilde{Q}^{\hat{0}\hat{l}} \nonumber \\
  & + \tfrac{2}{3} \lambda^2 D_{\tau}a^{\hat{i}} \delta_{\hat{j}\hat{i}} \delta_{\hat{k}\hat{l}} \partial_\tau \tilde{Q}^{\hat{0}\hat{l}} -  \lambda^2 a_{\hat{i}} a_{\hat{k}} \delta_{\hat{j}\hat{l}} \partial_\tau \tilde{Q}^{\hat{i}\hat{l}} + \lambda^2 a_{\hat{i}} a_{\hat{j}} \delta_{\hat{k}\hat{l}} \partial_\tau \tilde{Q}^{\hat{i}\hat{l}} -  \lambda^2 a_{\hat{i}} a_{\hat{k}} \delta_{\hat{j}\hat{l}} \partial_\tau \tilde{Q}^{\hat{l}\hat{i}} \nonumber \\
  & + \lambda^2 a_{\hat{i}} a_{\hat{j}} \delta_{\hat{k}\hat{l}} \partial_\tau \tilde{Q}^{\hat{l}\hat{i}} -  \tfrac{1}{3} \lambda^2 a_{\hat{i}} a^{\hat{i}} \delta_{\hat{j}\hat{l}} \delta_{\hat{k}\hat{m}} \partial_\tau \tilde{Q}^{\hat{l}\hat{m}} + \tfrac{1}{3} \lambda^2 a_{\hat{i}} a^{\hat{i}} \delta_{\hat{j}\hat{l}} \delta_{\hat{k}\hat{m}} \partial_\tau \tilde{Q}^{\hat{m}\hat{l}} + \tfrac{1}{3} \lambda^2 \delta_{\hat{j}\hat{i}} \delta_{\hat{k}\hat{l}} \partial_\tau{}^3 \tilde{Q}^{\hat{i}\hat{l}} \nonumber \\
  & -  \tfrac{1}{3} \lambda^2 \delta_{\hat{j}\hat{i}} \delta_{\hat{k}\hat{l}} \partial_\tau{}^3\tilde{Q}^{\hat{l}\hat{i}} + \mathcal{O}(\lambda^3)
\end{align}
Inserting covariant versions of these expressions into the first term on the RHS of Eq. (\ref{eq:diffconfirmation}) gives
\begin{subequations}
  \begin{align}
      \mathcal{P}^{\kappa}{}_\nu \int d^3 \Sigma_{\tilde{\mu}} m^{\tilde{\mu}} g^\nu{}_{\tilde{\lambda}}F_R^{\tilde{\lambda} \tilde{\rho}} j_{\tilde{\rho}} = \mathcal{P}^\kappa{}_\nu\bigg[&  \tfrac{2}{3} \lambda^2  \tilde{q}^2 D_{\tau}a^{\nu}  + \tfrac{2}{3} \lambda^3 \tilde{q} a^{\nu} D_{\tau}a^{\mu} \mathcal{P}_{\mu \lambda}  \tilde{Q}^{\rho \lambda} u_\rho -  \tfrac{2}{3} \lambda^3 \tilde{q} a^{\nu} D_{\tau}a_{\mu} \tilde{Q}^{\mu\rho} u_\rho    \nonumber \\
    & +  \tfrac{2}{3} \lambda^3 \tilde{q} D_{\tau}{}^2 a^{\nu}  \tilde{Q}^{\lambda \rho} u_\lambda u_\rho + \tfrac{2}{3} \lambda^3 \tilde{q} D_{\tau}a^{\nu}  a_{\mu} \tilde{Q}^{\rho \mu} u_\rho  +  \tfrac{2}{3} \lambda^3 \tilde{q} D_{\tau}a^{\nu} D_\tau \left( \tilde{Q}^{\lambda \rho} u_\lambda u_\rho \right) \notag\\&+ \tfrac{2}{3} \lambda^3 \tilde{q} a_{\mu} a^{\mu}  \delta_{\hat{k}\hat{j}}D_\tau\left(u_\lambda \mathcal{P}^\nu{}_\rho \tilde{Q}^{\lambda \rho}\right)  - \tfrac{2}{3} \lambda^3 \tilde{q} D_{\tau}a_{\sigma}  \mathcal{P}^\sigma{}_\mu D_\tau \left(\mathcal{P}^\nu{}_\lambda \mathcal{P}^\mu{}_\rho \tilde{Q}^{\lambda\rho}\right) \notag\\
        & +  \tfrac{2}{3} \lambda^3 \tilde{q} D_\tau\left(\mathcal{P}^\nu{}_\mu D_\tau\left(\mathcal{P}^\mu{}_\lambda D_\tau \left(\mathcal{P}^\lambda{}_\rho u_\sigma \tilde{Q}^{\sigma \rho} \right)\right)\right)\bigg] + \mathcal{O}(\lambda^4),\\
        u_\nu \int d^3 \Sigma_{\tilde{\mu}} m^{\tilde{\mu}} g^\nu{}_{\tilde{\lambda}}F_R^{\tilde{\lambda} \tilde{\rho}} j_{\tilde{\rho}} =& - \tfrac{2}{3} \lambda^3 \tilde{q} a_{\mu} D_{\tau}a^{\mu}  \tilde{Q}^{\nu \lambda} u_\nu u_\lambda  +  \tfrac{2}{3} \lambda^3 \tilde{q} D_{\tau}{}^2 a_{\mu} \mathcal{P}^\mu{}_\nu  \tilde{Q}^{\nu\lambda} u_\lambda  -  \tfrac{2}{3} \lambda^3 \tilde{q} a_{\nu} D_{\tau}a_{\mu} \tilde{Q}^{\nu\mu}  \nonumber \\
  &  +  \tfrac{4}{3} \lambda^3 \tilde{q} D_{\tau}a^{\mu} \mathcal{P}_{\mu \nu} D_\tau\left(\mathcal{P}^\nu{}_\lambda u_\rho \tilde{Q}^{(\rho\lambda)}\right)  + \mathcal{O}(\lambda^4)
  \end{align}
\end{subequations}
To evaluate the second term on the right hand side of
(\ref{eq:diffconfirmation}), we note from Eqs. (\ref{eq:momFlatDef}),
(\ref{eq:harteMomentum}),(\ref{eq:exactForce}), and (\ref{eq:scaledPdef}) that
it is given by the right hand sides of Eq.(\ref{eq:CovDMom2}), multiplied by
$\lambda$, and with the external fields set to zero. Equation
(\ref{eq:diffconfirmation}) thus evaluates to
\begin{align}
  D_\tau P_H^\kappa - D_\tau P_B^\kappa = \lambda D_\tau \bigg[&  \tfrac{2}{3} \lambda \tilde{q}^2 a^\kappa  - \tfrac{4}{3} \lambda^2 \tilde{q} a^{\kappa} \left( D_\tau \tilde{Q}^{\lambda \eta}u_\lambda u_\eta + \tfrac{2}{5}  a_{\nu} \tilde{Q}^{\lambda \nu} u_\lambda + a_{\nu} \tilde{Q}^{\nu \lambda} u_\lambda\right) - \tfrac{8}{3} \lambda^2 \tilde{q} a_{\nu}\mathcal{P}^\kappa{}_\mu D_\tau \tilde{Q}^{[\nu \mu]} \notag\\
  & +  \lambda^2 \tilde{q} a_{\nu} D_\tau \tilde{Q}^{\kappa \nu}  - \tfrac{2}{3} \lambda^2 u^\kappa a_\mu a_\nu \tilde{Q}^{\mu \nu} -  \tilde{q} \lambda^2 \mathcal{P}^\kappa{}_\mu\left( \tfrac{2}{3}D_\tau^2 \tilde{Q}^{\lambda \mu}u_\lambda + \tfrac{4}{15} a_\nu a^\nu \tilde{Q}^{\lambda \kappa}u_\lambda + \tfrac{2}{3} a_\nu a^\nu \tilde{Q}^{\kappa \lambda}u_\lambda\right) \notag\\
  &- \tfrac{2}{3} \lambda^2  (\mathcal{P}^\kappa{}_\mu + u^\kappa u_\mu) D_\tau a_\nu \tilde{Q}^{\nu \mu}\bigg] + \mathcal{O}(\lambda^4).
\end{align}
The right hand side is a total derivative as required, so our results satisfy the consistency condition.

\section{Conclusions}

In this paper, we have demonstrated the use of rigorous, limit based methods for
deriving higher-order self forces. Via an extension to the method first
introduced by GHW, combined with reasoning motivated by the work of Harte
\cite{HarteExact}, we have derived the entire self force effect through second
order without any ad hoc regularization.  These methods also yield the full
multipole dependence of radiation-reaction effects. The dipole dependence of the
first order radiation-reaction force was derived by GHW, and we find the
analogous second order dependence on dipole and quadrupole contributions. Our
results contain the first extended body dependence of any second order self
force, electromagnetic or otherwise, as well as the first explicit expression
for the self torque, which first arises at second order.

\section*{Acknowledgments}

We thank Abe Harte and Justin Vines for helpful conversations. This research was
supported in part by NSF grants PHY-1404105 and PHY-1707800.

 \appendix

\section{Convergence of integrals for bare spin and momentum} \label{app:sescaling} 

In this appendix, we show that the integral (\ref{eq:harteMomentum})
\begin{equation} \label{eq:harteMomentumapp}
  P_\tau(\xi) = \int_{\Sigma_\tau} T^{\tilde{\mu} \tilde{\nu}} \xi_{\tilde{\mu}} d\Sigma_{\tilde{\nu}}
\end{equation}
is well defined in Minkowski spacetime when $\xi^{\tilde{\mu}}$ is one of the
ten Killing vector fields, $\Sigma_\tau$ is a future null cone, and
$T^{\tilde{\mu} \tilde{\nu}}$ is the stress-energy tensor (\ref{eq:bodySE}) that
involves the retarded self-field. Different choices of Killing vector field
$\xi^{\tilde{\mu}}$ give rise to our definitions (\ref{eq:momSpinFlatDef}) of
linear momentum and spin.

We fix a point $z_\tau$ on the center of mass worldline and introduce coordinates $(u,r,\theta,\phi) = (u,r,\theta^1,\theta^2) = (u,r,\theta^A)$ such that the metric is
\begin{equation} \label{eq:flatmetric}
  ds^2 = -2 du dr - du^2 + r^2 d\Omega^2
\end{equation}
and that the null cone $\Sigma_\tau$ is the surface $u=\tau=\text{constant}$.
We define $n_\mu = -(du)_\alpha$, the null normal to $\Sigma_\tau$. The integral
(\ref{eq:harteMomentumapp}) can be written as

\begin{equation} \label{eq:nullintdependence}
  P_\tau(\xi) \propto \int_0^\infty dr r^2 \int d^2 \Omega Q_\mu \xi^\mu,
\end{equation}
where we have dropped the tildes for simplicity and
\begin{equation} \label{eq:projectedSE}
  Q_\mu = T_{\mu \nu} n^\nu.
\end{equation}

A priori, we would not expect the integral (\ref{eq:nullintdependence}) to
converge, since the leading order components of $T_{\mu \nu}$ scale as
$1/r^2$. However, we shall see that cancellations occur because the surface
$\Sigma_\tau$ is asymptotically a surface of constant phase for the outgoing
radiation. From Eq.(\ref{eq:nullintdependence}), a sufficient condition for
convergence is that
\begin{equation} \label{eq:convcondition}
  \int d^2 \Omega Q_\mu \xi^\mu = \mathcal{O}(r^{-4})
\end{equation}
as $r\rightarrow\infty$.

The general form of a Killing vector field in the coordinates
(\ref{eq:flatmetric}) as $r\rightarrow\infty$ is \cite{flanagan15}
\begin{align} \label{eq:asymptkilling}
 \vec{\xi} =& \left[\alpha + \tfrac{1}{2} u \Psi + \mathcal{O}(r^{-1})\right]\partial_u + \left[Y^A + \mathcal{O}(r^{-1})\right] \partial_A - \left[\tfrac{1}{2} r \Psi + \mathcal{O}(1)\right] \partial_{r},
\end{align}
where $Y^A(\theta^B)$ is a conformal Killing vector field on the 2-sphere that
encodes rotation and boosts, $\Psi=D_A Y^A$, and $D_A$ is the covariant
derivative operator with respect to the 2-sphere metric $h_{A B}$ defined by
$d\Omega^2 = h_{A B} d\theta^A d\theta^B$. The function $\alpha(\theta^B)$ is a
linear combination of $l=0$ and $l=1$ spherical harmonics and encodes
translations.

Now inserting (\ref{eq:asymptkilling}) into (\ref{eq:convcondition}), we find
the sufficient condition for convergence is
\begin{align} \label{eq:integralScalings}
  \int d^2 \Omega \bigg\{&\left[\tfrac{1}{2}u \Psi + \alpha + \mathcal{O}(r^{-1})\right] Q_u + \left[Y^A + \mathcal{O}(r^{-1})\right] Q_A + \left[-\tfrac{1}{2} r \Psi + \mathcal{O}(1)\right]Q_r\bigg\} = \mathcal{O}(r^{-4}),
\end{align}
which will be satisfied if
\begin{subequations} \label{eq:individualScalings}
  \begin{align}
    Q_u = \mathcal{O}(r^{-4}),\\
    Q_A = \mathcal{O}(r^{-4}),\\
    Q_r = \mathcal{O}(r^{-5}).
  \end{align}
\end{subequations}

Consider first the scalar case. When the scalar charge density $\rho$ is smooth,
the method of Sec. 11.1 of \cite{waldbook} can be used to show that the retarded
scalar field $\Phi^{(\text{self})}$ has an expansion near future null infinity
of the form
\begin{equation} 
\Phi^{(\text{self})} = \frac{f(u,\theta^A)}{r} + \frac{g(u,\theta^A)}{r^2} + \mathcal{O}(r^{-3}),
\end{equation}
for some smooth functions $f$ and $g$. Inserting this expansion into
Eqs.(\ref{eq:scalarfieldSE}),(\ref{eq:SEselfcrossext}),(\ref{eq:bodySE}), and
(\ref{eq:projectedSE}) yields
\begin{subequations} \label{eq:scalarSEscalings}
  \begin{align}
    Q_r =& \frac{-f^2}{r^4} + \mathcal{O}(r^{-5})\\
    Q_u =& \frac{-1}{2 r^4} \left[f^2 + h^{A B} D_A f D_B f\right] + \mathcal{O}(r^{-4})\\
    Q_A =& \frac{1}{r^3} f D_A f + \mathcal{O}(r^{-4})
  \end{align}
\end{subequations}
It can be seen that these expressions do not satisfy the scalings (\ref{eq:individualScalings}). However, inserting the expressions (\ref{eq:scalarSEscalings}) into (\ref{eq:integralScalings}) and integrating by parts on the two-sphere, we find that the leading order terms cancel and so the condition (\ref{eq:integralScalings}) is satisfied.

Turn now to the electromagnetic case. We can use the method of Sec 11.1 of \cite{waldbook} to deduce the asymptotic scaling of the component of the retarded field $F_{\mu \nu}^{(\text{self})}$. Defining $\rho=r^{-1}$, the metric can be written as $ds^2 = \rho^{-2} d\tilde{s}^2$ with
\begin{equation} \label{eq:rescaledMetric}
  d\tilde{s}^2 = - \rho^2 du^2 - 2 du d\rho + d\Omega^2.
\end{equation}
Since the field equations (\ref{eq:maxwell}) are conformally invariant away from sources, $F_{\mu \nu}^{(\text{self})}$ is a solution of the equations in the metric (\ref{eq:rescaledMetric}) and hence is a smooth function of $(\rho, u, \theta^A)$ at $\rho=0$, i.e. on future null infinity. It follows that for general solutions with smooth sources
\begin{subequations} \label{eq:maxwellscalings}
  \begin{align}
  F_{u r}^{(\text{self})} =& \mathcal{O}(r^{-2}),\\
  F_{u A}^{(\text{self})} =& \mathcal{O}(1),\\
  F_{r A}^{(\text{self})} =& \mathcal{O}(r^{-2}),\\
  F_{A B}^{(\text{self})} =& \mathcal{O}(1)
  \end{align}
\end{subequations}
as $r\rightarrow\infty$. From Eqs. (\ref{eq:fieldSE}),(\ref{eq:SEselfcrossext}),(\ref{eq:bodySE}), and (\ref{eq:projectedSE}) we find that
\begin{subequations}\label{eq:projectedSEmaxwell}
  \begin{align} 
    Q_r =& -\frac{1}{r^2} F_{rA}^{(\text{self})} F_{r B}^{(\text{self})} h^{A B},\\
    Q_A =& - F_{A r}^{(\text{self})} F_{u r}^{\text{self}} - \frac{1}{r^2} F_{A B}^{(\text{self})} F_{r C}^{(\text{self})} h^{BC},\\
    Q_u =& -\frac{1}{2} F_{u r}^{(\text{self})} {}^2 - \frac{1}{2 r^2} F_{rA}^{(\text{self})} F_{r B}^{(\text{self})} h^{A B} - \frac{1}{r^4} F_{AB}^{(\text{self})} F_{CD}^{(\text{self})} h^{AC} h^{B D}.
  \end{align}
\end{subequations}
Inserting the scalings (\ref{eq:maxwellscalings}) into the expressions (\ref{eq:projectedSEmaxwell}) we find that the conditions for convergence (\ref{eq:individualScalings}) are satisfied.

\section{Scalar laws of motion} \label{app:ScalarLaws}

\subsection{Renormalized scalar moments}

As for the electromagnetic case, we find it useful to introduce a renormalized
set of moments to describe the scalar charge distribution, modifying the
rescaled moments $\tilde{q_S}$,$\tilde{Q}_S^{\mu}$, and $\tilde{Q}_S{}^{\mu
  \nu}$ given in Eq. (\ref{eq:baremultScalar}). Unlike the electromagnetic case,
the scalar charge and so may be renormalized \footnote{That is, the definition
  of the charge depends on the choice of hypersurface, so it is natural to allow
  a redefinition of the charge in order to simplify the equations of motion.},
so possesses an ambiguity in the chargelike degrees of freedom. The
renormalized charge is
\begin{align}
    q_{S} = \tilde{q}_{S} + \lambda D_\tau \tilde{Q}_{S}{}^{\mu}  u_\mu - \lambda^2 D_\tau \left(u_\mu \tilde{Q}_{S}{}^{\mu \nu} a_\nu\right)  + \mathcal{O}(\lambda^3).
\end{align}
The renormalized projected dipole is
\begin{align}
  Q_S{}^\mu = \mathcal{P}^\mu{}_\nu \left(\tilde{Q}_S{}^\nu +\lambda D_\tau \tilde{Q}_S^{\nu \lambda} u_\lambda\right) + \mathcal{O}(\lambda^2),
\end{align}
which is explicitly orthogonal to the 4-velocity.  We define
the renormalized projected quadrupole as
\begin{align}
Q_S{}^{\mu \nu} = \mathcal{P}^\mu{}_\lambda \mathcal{P}^\nu{}_\sigma \left(\tilde{Q}_S{}^{\lambda \sigma }\right) + \mathcal{O}(\lambda),
\end{align}
which is explicitly orthogonal to $u^\mu$ in both of its indices, $u_\mu Q_{S}^{\mu \nu} = u_\nu Q_{S}^{\mu \nu} = 0$.

In addition, as in the electromagnetic case, we find it useful to define a renormalized mass and a renormalized spin. The definitions are
\begin{subequations}
  \begin{align}
      m + u_\mu \tilde{P}^{\mu} =& - \lambda u_\nu  \Phi^{(\text{ext});\nu} \tilde{Q}_S^{\mu} u_\mu + \lambda\tilde{q} D_\tau \tilde{q} -  \lambda^2 u_\lambda \Phi^{(\text{ext});\lambda}{}_{\mu} \mathcal{P}^\mu{}_\nu \tilde{Q}_S^{\nu\rho} u_\rho \notag\\&+ \lambda^2 u_\mu \Phi^{(\text{ext});\mu} a_{\nu} \tilde{Q}_S^{\nu \lambda} u_\lambda     + \tfrac{1}{3}\lambda^2 \tilde{q} a_{\mu} a^{\mu} \tilde{Q}_S^{\nu} u_\nu + \tfrac{1}{3} \lambda^2 \tilde{q} a_{\nu}  D_{\tau}\tilde{Q}_S^{\nu} \notag\\&+  \lambda^2 \tilde{q} D_{\tau}{}^2 \tilde{Q}_S^{\mu} u_\mu     -  \tfrac{2}{3} \lambda^2 a_{\mu} \tilde{Q}_S^{\mu} D_\tau \tilde{q}   + \lambda^2 D_{\tau}\left(\tilde{Q}_S^{\mu} u_\mu\right) D_\tau \tilde{q} + \mathcal{O}(\lambda^3)\\
  S{}^{\mu \nu} =& \tilde{S}^{\mu \nu}  + 2 \lambda \Phi^{(\text{ext});[\mu} \tilde{Q}_S^{\nu] \lambda} u_\lambda+ \tfrac{2}{3} \lambda \tilde{q} a^{[\mu} \tilde{Q}_S^{\nu]} + \tfrac{2}{3} \lambda u^{[\nu} D_\tau \left(\tilde{q} \tilde{Q}_S^{\mu]}\right)  + \mathcal{O}(\lambda^2).
  \end{align}
\end{subequations}
\subsection{Scalar self force in terms of renormalized moments}

As in the electromagnetic presentation, we decompose the self force and rest mass evolution as
\begin{subequations}
  \begin{align}
    m a^{\mu} =& f_S^{(0) \mu} + \lambda f_S^{(1) \mu} + \lambda^2 f_S^{(2) \mu} + \mathcal{O}(\lambda^3)\\
    D_\tau m =& \mathcal{F}_S^{(0)} + \lambda \mathcal{F}_S^{(1)} + \lambda^2 \mathcal{F}_S^{(2)} + \mathcal{O}(\lambda^3)
  \end{align}
\end{subequations}

Following similar steps to the electromagnetic derivation, we find the leading force and mass evolution
\begin{subequations}
  \begin{align}
    f^{(0)\mu}_S =& q_S \mathcal{P}^\sigma{}_\kappa  \Phi^{(\text{ext})\kappa}\\
    \mathcal{F}^{(0)}_S =&-q_S \Phi^{(\text{ext})\mu}u_\mu,
  \end{align}
\end{subequations}
where $\Phi^{(\text{ext})\mu} \equiv \nabla^\mu \Phi^{(\text{ext})}$.  The
GHW-order scalar self force and mass evolution, 
\begin{subequations}
  \begin{align}
       f_S^{(1) \mu} =&\, \mathcal{P}^\sigma{}_\kappa\bigg[ q_S \Phi^{(\text{ext})\kappa} +   \Phi^{(\text{ext})\kappa}{}_{;\mu}Q_S{}^{\mu} +  \tfrac{1}{3} D_{\tau}a^{\kappa} q_S^2  + a^{\kappa} q_S D_\tau q_S  - 2  D_\tau \left(Q_S{}^{[\kappa}  \Phi^{(\text{ext})\mu]}u_\mu \right) + D_{\tau}\left(a_{\mu} S^{\kappa\mu}\right)\bigg] \\
    \mathcal{F}^{(1)}  =&\, -q_S \Phi^{(\text{ext})\mu}u_\mu - u_\nu \Phi^{(\text{ext})\nu}{}_{;\mu } Q_S{}^{\mu} - 2  \Phi^{(\text{ext})\mu}u_\mu a_{\nu} Q_S^{\nu}    + q_S  D_\tau{}^2 q_S
  \end{align}
\end{subequations}

These results are new except for the monopole terms, which can be found in
\cite{quinn}.  The second-order results can be expressed as a sum of as a sum of
dipole and quadrupole contributions:

\begin{subequations}
  \begin{align}
    f_{S}^{(2) \mu} =&\,     f_{S}^{(2) \mu}{}_{\text{dipole}} + f_{S}^{(2) \mu}{}_{\text{quadrupole}},\notag\\
   \mathcal{F}_S^{(2)} =&\, \mathcal{F}_S^{(2)}{}_{\text{dipole}} + \mathcal{F}_S^{(2)}{}_{\text{quadrupole}},
  \end{align}
\end{subequations}
As for the electromagnetic case, there are no explicit monopole terms at this
order. The explicit, new, dipole and quadrupole contributions to the self force
are:
\begin{subequations}
  \begin{align}
    f_{S}^{(2) \mu}{}_{\text{dipole}}= \mathcal{P}^\sigma{}_\kappa \bigg[&  -  \tfrac{1}{3} q_{S}{}  a^{\kappa} D_{\tau}Q_{S}{}^{\nu}  a_{\nu} -  \tfrac{1}{3} a_{\nu} a^{\nu} \left(q_{S}{} D_{\tau}Q_{S}{}^{\kappa}  - D_\tau q_{S}{} Q_{S}{}^{\kappa}\right)  -  \tfrac{2}{3} q_{S}{} D_{\tau}a^{\kappa}  a_{\mu}   Q_{S}{}^{\mu} \nonumber \\
  &   -  \tfrac{1}{3} Q_{S}{}^{\kappa} D_\tau{}^3 q_{S}{} + \tfrac{1}{3}  q_{S}{} D_{\tau}{}^3 Q_{S}{}^{\kappa} -  q_{S}{} a^{\kappa} D_{\tau}a_{\mu}  Q_{S}{}^{\mu}  - D_\tau \left(D_\tau q_{S}{} D_{\tau}{}Q_{S}{}^{\kappa} \right) \bigg], \\
    f_{S}^{(2) \mu}{}_{\text{quadrupole}} = \mathcal{P}^\sigma{}_\kappa\bigg[& \tfrac{1}{2}  \nabla \Phi^{(\text{ext})\kappa}{}_{;\mu \nu} Q_{S}{}{}^{\mu \nu} + \tfrac{1}{2} Q_{S}{}{}^{\rho}{}_\rho D_\tau{}^2 \Phi^{(\text{ext})\kappa}   - D_{\tau}\left(u_\mu  \Phi^{(\text{ext})\mu}{}_{;\nu} Q_{S}{}{}^{\kappa\nu} \right) \notag\\& + \Phi^{(\text{ext})\kappa}{}_{;\mu}  Q_{S}{}{}^{\mu\nu} a_{\nu} -  D_{\tau}\left(\Phi^{(\text{ext})\mu} u_\mu   Q_{S}{}{}^{\kappa\nu} a_{\nu}\right)     + \tfrac{1}{2} \Phi^{(\text{ext})\kappa}{}_{;\mu} a^{\mu} Q_{S}{}{}^{\rho}{}_\rho    \nonumber \\
     & -  D_\tau \Phi^{(\text{ext})\mu} u_\mu Q_{S}{}{}^{\kappa\nu}  a_{\nu}    +D_\tau \left( D_{\tau}{}Q_{S}{}{}^{\rho}{}_\rho \Phi^{(\text{ext})\kappa}\right)\notag\\&  - 2 \Phi^{(\text{ext})\kappa} a_{\mu} a_{\nu}  Q_{S}{}^{\mu \nu} +  a^{\kappa} \Phi^{(\text{ext})\mu}u_\mu D_{\tau} Q_{S}{}{}^{\rho}{}_\rho \bigg],
  \end{align}
\end{subequations}
and the explicit, new, dipole and quadrupole contributions to the mass evolution are
\begin{subequations}
  \begin{align}
    \mathcal{F}_S{}_{\text{dipole}}^{(2)} =&  \tfrac{1}{3} q_{S}{}  D_{\tau}a_{\mu} \mathcal{P}^\mu{}_\nu D_{\tau}Q_{S}{}{}^{\nu}  -  a_{\mu}D_\tau \left( D_\tau q_{S}{} Q_{S}{}{}^{\mu} \right) -  \tfrac{4}{3} D_\tau q_{S}{} D_{\tau}a_{\mu} Q_{S}{}{}^{\mu}, \\
      \mathcal{F}_S{}_{\text{quadrupole}}^{(2)} =&  - \Phi^{(\text{ext})\mu} u_\mu a_{\nu} a_{\lambda} Q_{S}{}{}^{\nu \lambda} - 2 u_\lambda\Phi^{(\text{ext})\lambda}{}_{;\mu} Q_{S}{}{}^{\mu \nu} a_{\nu}  - \tfrac{1}{2}  u_\mu \Phi^{(\text{ext})\mu }{}_{;\nu \lambda} Q_{S}{}{}^{\nu \lambda}\nonumber \\
  &- \tfrac{1}{2} u_\mu \Phi^{(\text{ext})\mu}{}_{;\nu \lambda} u^\nu u^\lambda Q_{S}{}{}^{\rho}{}_\rho  + a_{\mu} D_{\tau}\left(\Phi^{(\text{ext})\mu} Q_{S}{}{}^{\rho}{}_\rho \right).
  \end{align}
\end{subequations}

\subsection{Scalar self torque}

The self torque of a scalar charged body in terms of the renormalized moments is,
\begin{align}
    D_\tau S{}^{\kappa \lambda} \mathcal{P}_\kappa{}^\sigma \mathcal{P}_\lambda{}^\rho =& \mathcal{P}^\sigma{}_\kappa  \mathcal{P}^\rho{}_\lambda \bigg[ 2\Phi^{(\text{ext})[\kappa} Q_S{}^{\lambda]}  +  2\lambda  \Phi^{(\text{ext})[\kappa}{}_{;\mu} Q_S{}^{\lambda]\mu} \notag\\&\hspace{1.5cm}+  \tfrac{2}{3} \lambda q_S D_{\tau}a^{[\kappa}  Q_S{}^{\lambda]}  + 2 \lambda  \Phi^{(\text{ext})[\kappa} Q_S{}^{\lambda]\mu} a_{\mu}   \bigg] + \mathcal{O}(\lambda^2)
\end{align}

\subsection{Scalar point particle reduced order}

We again specialize to a monopole body, for which $S_{\mu \nu} = 0$,$Q_S^{\mu}
=0$, $Q_S^{\mu \nu} =0$, and present the reduced order equation of motion.

Here we give the acceleration and rest mass evolution of the point-particle
limit for a scalar charge, similar to the expressions for an electromagnetic
charge given in Sec. \ref{sec:empointparticle}. Note that the lack of a conserved total charge
for the scalar case makes this limit somewhat arbitrary - we take it to indicate
the vanishing of all moments of the body apart from the renormalized charge $q_{S}$. 

The acceleration, in terms of only the external field and the charge, is
\begin{align}
    a^{\kappa} =& \frac{q_S}{m}\mathcal{P}^\kappa{}_\sigma \Phi^{(\text{ext}) \sigma} + \frac{4}{3} \lambda \frac{q_S{}^2}{m{}^2} D_\tau q_S \mathcal{P}^\kappa{}_\sigma \Phi^{(\text{ext}) \sigma} + \frac{1}{3} \lambda \frac{q_S{}^3}{m^2} \mathcal{P}^\kappa{}_\sigma u^\mu \Phi^{(\text{ext})\sigma}{}_{;\mu} \notag\\
  & + \frac{20}{9} \lambda^2 \frac{q_S{}^3}{m{}^3} \left(D_\tau q_S\right)^2 \mathcal{P}^{\kappa}{}_\sigma \Phi^{(\text{ext})\sigma} + \lambda^2 \frac{q_S{}^4}{m{}^3} \mathcal{P}^\kappa{}_\sigma \left(\frac{10}{9} D_\tau q_S \Phi^{(\text{ext}) \sigma}{}_{;\mu} u^\mu + \frac{4}{9} D_\tau{}^2 q_S \Phi^{(\text{ext})\sigma} \right)\notag\\
  &+ \frac{1}{9}  \lambda^2\frac{q_S{}^5}{m{}^3} \mathcal{P}^\kappa{}_\lambda \Phi^\lambda{}_{;\mu \nu} u^\mu u^\nu - \frac{4}{9} \lambda^2 \frac{q_S^5}{m^4} D_\tau q_S \mathcal{P}^\kappa{}_\sigma \Phi^{(\text{ext}) \sigma} u_\lambda \Phi^{(\text{ext})\lambda} \notag\\
  &+ \frac{1}{9}  \lambda^2\frac{q_S{}^6}{m^4} \left(- \mathcal{P}^\kappa{}_\lambda \Phi^{(\text{ext})\lambda}{}_{;\mu} u^\mu u_\sigma \Phi^{(\text{ext})\sigma} + \mathcal{P}^\kappa{}_\lambda \Phi^{(\text{ext})\lambda} \Phi^{(\text{ext})}{}_{;\mu \nu} u^\mu u^\nu  + \mathcal{P}^{\kappa}{}_\sigma \Phi^{(\text{ext}) \sigma}{}_{;\mu}  \Phi^{(\text{ext})\mu}\right) + \mathcal{O}\left(\lambda^3\right)
\end{align}
The  evolution of the renormalized mass, in terms of only the external field and the charge, is simply
\begin{align}
  D_\tau m = q_S \Phi^{(\text{ext})\mu}u_\mu   +  \lambda q_S  D_\tau{}^2 q_S + \mathcal{O}(\lambda^3)
\end{align}

\bibliographystyle{apsrev4-1}
\bibliography{2ndOrdSelfForce}

\end{document}